\documentclass{aa} % for the letters 
\usepackage{txfonts}
\usepackage{graphicx}
\usepackage{subfig}
\usepackage{caption}
\usepackage{natbib}
\usepackage{multirow}
\usepackage{float}
\title{The solar wind from a stellar perspective:}
\subtitle{how do low-resolution data impact the determination of wind properties?}
\author{S. Boro Saikia \inst{1}, M. Jin \inst{2,3}, C.P. Johnstone \inst{1}, T. L\"uftinger \inst{1}, M. G\"udel \inst{1}, V. S. Airapetian \inst{4,5}, K. G. Kislyakova \inst{1}, and C. P. Folsom \inst{6}}
 \institute{University of Vienna, Department of Astrophysics,
              T\"urkenschanzstrasse 17, 1180 Vienna, Austria \label{inst1} \and
Lockheed Martin Solar and Astrophysics Laboratory, Palo Alto, CA 94304, USA \label{inst2} \and
SETI institute, Mountain View, CA 94043, USA \label{inst3}\and
Sellers Exoplanetary Environments Collaboration, NASA Goddard Space Flight Center, Greenbelt, USA \label{inst4}\and
American University, Washington DC, USA \label{inst5}\and
IRAP, Universit\'{e} de Toulouse, CNRS, UPS, CNES, 14 Avenue Edouard Belin, 31400, Toulouse, France \label{inst6}
} 
\titlerunning{solar wind}
\authorrunning{boro saikia}
\abstract{Due to the effects that they can have on the atmospheres of exoplanets, stellar winds have recently received 
significant attention in the literature. Alfv\'en-wave-driven 3D magnetohydrodynamic (MHD) models, which are increasingly 
used to predict stellar wind properties, contain unconstrained parameters and rely on low-resolution stellar magnetograms.
}{In this paper, we explore the effects of the input Alfv\'en wave energy flux and the surface magnetogram on the wind properties 
predicted by the Alfv\'en Wave Solar Model (AWSoM) model for both the solar and stellar winds.
}{We lowered the resolution of two solar magnetograms during solar cycle maximum and minimum using
spherical harmonic decomposition. The Alfv\'en wave energy was altered based on non-thermal velocities determined from a far ultraviolet (FUV) spectrum of 
the solar twin 18 Sco. Additionally, low-resolution magnetograms of three solar analogues, 18 Sco, HD 76151, and HN Peg, were obtained 
using Zeeman Doppler imaging (ZDI) and used as a proxy for the solar magnetogram. 
Finally, the simulated wind properties were compared to Advanced Composition Explorer (ACE) observations.}{AWSoM simulations using well constrained 
input parameters taken from solar observations can reproduce the observed solar wind mass loss and angular momentum loss rates.
The simulated wind velocity, proton density, and ram pressure differ from ACE observations by 
a factor of approximately two. The resolution of the magnetogram has a small impact on the wind properties and only during cycle maximum. 
However, variation in Alfv\'en wave energy influences the wind properties irrespective of the solar cycle activity level.
Furthermore, solar wind simulations carried out using the low-resolution magnetogram of the three
stars instead of the solar magnetogram could lead to an order of a magnitude difference in the simulated solar wind properties.}
{{The choice in Alfv\'en energy has a stronger influence on the wind output compared to the magnetogram resolution. The influence 
could be even stronger for stars whose input boundary conditions are not as well constrained as those of 
the Sun.}
Unsurprisingly, replacing the solar magnetogram with a stellar magnetogram 
could lead to completely inaccurate solar wind properties, and should be avoided in solar and stellar wind simulations. 
Further observational and theoretical work is
needed to fully understand the complexity of solar and stellar winds. }{}
\begin{document}

\maketitle
\section{Introduction}
Stellar magnetic fields are responsible for a large number of phenomena, 
including the emission of high-energy radiation and the formation of supersonic ionised winds. 
By driving atmospheric processes such as non-thermal losses to space, these winds play an 
important role in the evolution of planetary atmospheres and habitability \citep{tian08,kislyakova14,airapetian17}. 
As an example, the strong solar wind of the young Sun \citep{johnstone15b,airapetian16} 
in combination with the weaker magnetic field of early Earth  \citep{Tarduno10} 
led to higher compression of the Earth's magnetosphere. This resulted in wider opening of polar ovals 
and higher atmospheric escape rates than at present \citep{Airapetian16b}. 
It has been shown that planetary atmospheric loss in 
planets with a magnetosphere depends on the interplay between the solar 
wind strength, wind capture area of the planetary magnetosphere, 
and the  ability of the magnetosphere to recapture the atmospheric outflow, although 
the effect of magnetospheric compression on atmospheric loss rates are 
currently up for debate \citep{blackman18}.
For planets lacking any intrinsic magnetic field, 
the incoming stellar wind interacts directly with the atmosphere, leading to atmospheric escape 
through the plasma wake and from a boundary layer of the induced magnetosphere \citep{Barabash07,Lundin11}. Venus-like 
CO$_\mathrm{2}$-rich atmospheres are less prone to expansion and escape, but they are still sensitive to 
enhanced X-ray and extreme ultraviolet (XUV) fluxes, and wind erosion \citep{Lichtenegger10,Johnstone18}.
The same can be true for exoplanets orbiting young stars with stronger stellar winds leading to efficient 
escape of the atmosphere to space {\citep{wood02,Lundin11}}. It is therefore important to investigate stellar wind properties in Sun-like stars 
to understand their impact on habitability and also provide constraints on planetary atmospheres.

Observations of the solar wind taken by satellites such as the Advanced Composition Explorer \citep[ACE,][]{stone98,mccomas98} 
and \textit{Ulysses} \citep{mccomas03} have greatly improved our knowledge and understanding of the solar wind properties.
The solar wind can be broken down 
into the fast and the slow wind with median wind speeds of approximately 760~km~s$^{-1}$ and 400~km~s$^{-1}$ 
respectively \citep{mccomas03,johnstone15a}. The fast component arises from coronal holes and the slow component is launched from areas above
closed field lines, and from the boundary regions of open and closed field lines \citep{krieger73}. As the magnetic geometry
of the Sun changes during the solar cycle, the locations of the fast and slow components change 
without any considerable changes in their properties, such as speed or mass flux.
\emph{In situ} measurements by spacecrafts such as 
\textit{Ulysses} and \textit{Voyager} have found that the mass loss rate of the solar wind is 
$\sim$2$\mathrm\times$10$^{-14}$~$\mathrm{M_\sun}$~yr$^{-1}$, and that it  changes by a factor
of only two over the solar cycle \citep{cohen11}. Angular momentum loss rates vary 
by 30-40$\%$ over the solar cycle as shown by \citet{finley18}.  
This shows that despite the dramatic change in the surface magnetic field of the Sun during the 
cycle, the changes in the solar wind properties are not drastic.

Unfortunately, direct measurements of the properties of {low-mass} stellar winds are not available; instead techniques 
to indirectly measure stellar winds must be used, including reconstructing astrospheric Ly-$\alpha$ 
absorption \citep{wood01} and fitting rotational evolution models to observational constraints \citep{matt15,johnstone15b}. 
This is problematic for stellar wind modelling since we can neither constrain the model free parameters nor test our 
results observationally.
Attempts have been made to detect radio free-free emissions due to the presence of stellar winds in Sun-like stars 
\citep{drake93,vanoord97,gaidos00,villadsen14,fichtinger17}. Unfortunately there has been no detection so far but radio
observations have provided important upper limits on the wind mass loss rates of a handful of Sun-like stars. X-ray emission due to charge 
exchange between ionised stellar winds and  the neutral interstellar hydrogen have also been used to provide
upper limits on the mass loss rate due to stellar winds \citep{wargelin02}. For a limited sample of close-in transiting 
hot Jupiters, Lyman-$\alpha$ observations have been used to estimate the properties of the 
wind of the host star \citep{kislyakova14b,vidotto17}. The indirect method of astrospheric Lyman-$\alpha$
measurements \citep{wood01,wood04,wood05} is the only technique that has provided observed wind mass loss 
rates for some nearby Sun-like stars. Using this method \citet{wood05} showed that the mass loss rate has a power-law relation with magnetic activity, implying that more active stars have higher mass loss rates.  
Some stars do not appear to follow this trend and this method can only be applied to nearby stars that
are surrounded by at least partially neutral interstellar medium. 
We are therefore heavily dependent on wind models to enhance our understanding of stellar wind properties. 

Solar and stellar wind modelling faces multiple challenges, as we still lack {a complete} understanding of the heating, 
acceleration, and propagation of the wind. The outward acceleration of the wind takes place in 
large part due to thermal pressure gradients driven by the very large temperatures of 
coronal gas \citep{parker58}. However, measurements of the gas temperatures inside coronal holes show that the temperatures are not high enough to accelerate the wind to the speed of the fast 
component, and therefore another acceleration mechanism is required {\citep{cranmer09}}. 
The source of the wind heating and the nature of this additional acceleration mechanism are 
currently poorly understood \citep{cranmer19}. Alfv\'en waves {are} considered to be a 
likely key mechanism for solar wind heating and acceleration.
Observations taken using \textit{Hinode} \citep{kosugi07} and the Solar Dynamics Observatory \citep[SDO,][]{pesnell12} 
have shown that Alfv\'enic waves in the solar chromosphere have much stronger amplitudes
compared to their coronal counterpart \citep{depontieu07,mcintosh11}. The weakening of the waves as they reach the corona is 
attributed to the wave dissipation. The waves reflected by density and magnetic pressure 
gradients interact with the forward propagating 
waves resulting in wave dissipation, which in turn heats the lower corona. This provides the necessary energy
to propagate and accelerate the wind so that it can escape from the gravity of the star \citep{matthaeus99}.
It has been suggested that for very rapidly rotating stars, magneto-centrifugal forces also provide an 
important wind acceleration mechanism \citep{johnstone17}.

To tackle the wind heating problem, it has been common in solar and stellar wind models to assume a 
polytropic equation of state \citep{parker65,vanderholst07,cohen07,johnstone15a}, which states that the pressure, $p$, 
is related to the density, $\rho$, by \mbox{$p \propto \rho^\alpha$}, where $\alpha$ is the 
polytropic index and is typically taken to be \mbox{$\alpha \sim 1.1$}. 
This leads to the wind being heated implicitly as it expands. 
Free parameters in these models are the density and temperature at the base of the wind and the 
value of $\alpha$, all of which can be constrained for the solar wind from \emph{in situ} 
measurements \citep{johnstone15a}. However, these parameters are unconstrained for the winds of other stars.
An alternative is to use solar and stellar wind models that incorporate Alfv\'en waves, 
which are {becoming increasingly popular} \citep{cranmer11,suzuki13}. 
Some of the earliest Alfv\'en-wave-driven models date back to  
\citet{belcher71}, and \citet{alazraki71}. Multiple groups have developed 1D \citep{suzuki06,cranmer07}, 
2D \citep{usmanov00,matsumoto12}, and 3D \citep{sokolov13,vanderholst14,usmanov18} 
Alfv\'en-wave-driven solar wind models that can successfully simulate the current solar wind mass loss rates. 
In this work the 3D magnetohydrodynamic (MHD) model Alfv\'en Wave Solar 
Model (AWSoM; \citealt{vanderholst14}) is used, where Alfv\'en wave propagation 
and partial reflection leads to a {{turbulent cascade, heating and accelerating} the wind.  
The Alfv\'en wave energy is introduced using 
an input Alfv\'en wave Poynting flux ratio ($S_\mathrm{A}/B$, 
where $B$ is the magnetic field strength at the inner boundary of the simulation). For the Sun, 
\citet{sokolov13} established $S_\mathrm{A}/B$ to be 1.1~$\times$~10$^6$~Wm$^{-2}$~T$^{-1}$.
In stellar wind models, $S_\mathrm{A}/B$ is modified using scaling laws between the X-ray activity 
and magnetic field $B$  of the star \citep{pevtsov03,garraffo16,dong18}, which requires prior information about the former parameter. As stellar X-ray activity is known to exhibit variations, this approach will also lead to 
variation in $S_\mathrm{A}/B$ for a magnetically variable star. In the solar
case, the Poynting flux is well constrained from observations and nearly constant for all solar simulations.
The value of $S_\mathrm{A}/B$ is an important input parameter, but how a change in 
the Poynting flux ratio quantitatively changes the final wind output remains unknown. It is important 
to understand the relationship between $S_\mathrm{A}/B$ and the coronal and
wind properties, as often the Poynting flux ratio is a difficult parameter to directly determine from
stellar observations. 
\begin{figure*}
\centering
\captionsetup[subfigure]{labelformat=empty}
\subfloat[]{\includegraphics[width=.42\textwidth]{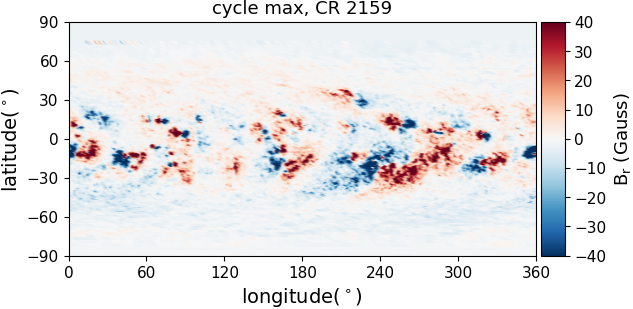}}\hspace{5mm}
\subfloat[]{\includegraphics[width=.42\textwidth]{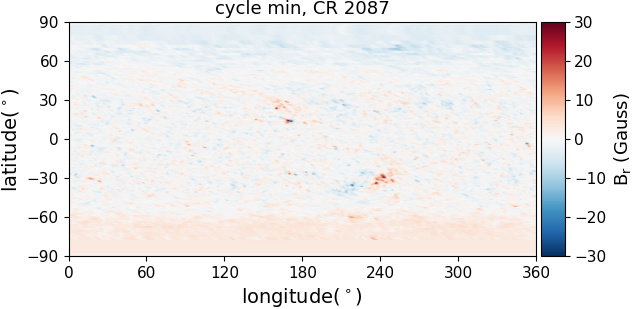}}\\[-2ex]
\subfloat[]{\includegraphics[width=.42\textwidth]{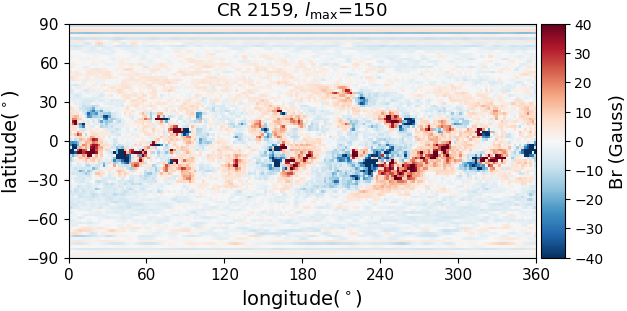}}\hspace{5mm}
\subfloat[]{\includegraphics[width=.42\textwidth]{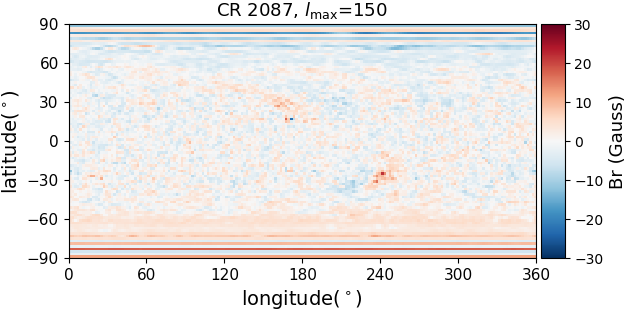}}\\[-2ex]
\subfloat[]{\includegraphics[width=.42\textwidth]{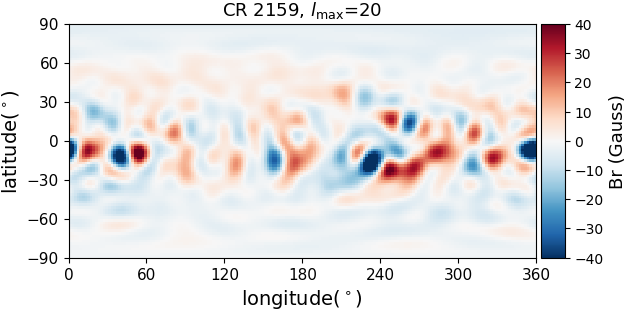}}\hspace{5mm}
\subfloat[]{\includegraphics[width=.42\textwidth]{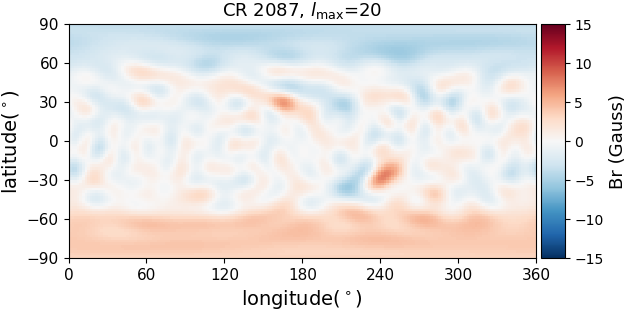}}\\[-2ex]
\subfloat[]{\includegraphics[width=.42\textwidth]{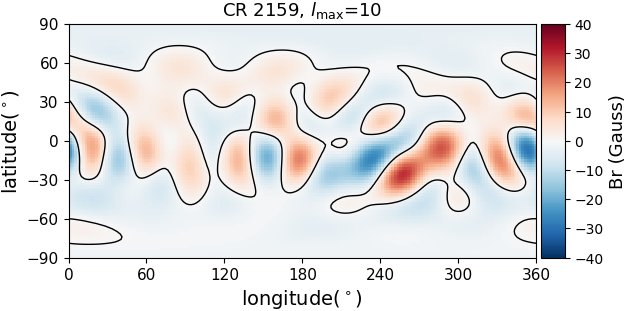}}\hspace{5mm}
\subfloat[]{\includegraphics[width=.42\textwidth]{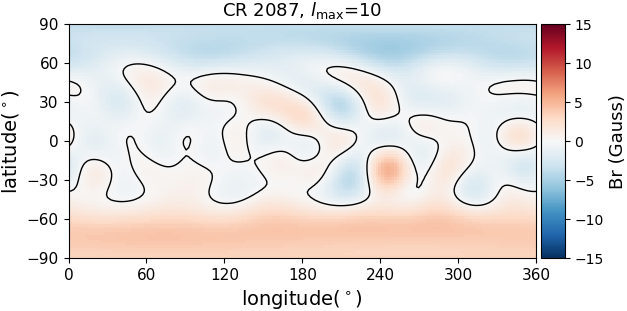}}\\[-2ex]
\subfloat[]{\includegraphics[width=.42\textwidth]{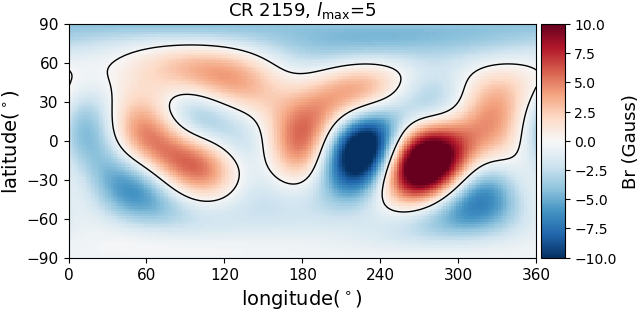}}\hspace{5mm}
\subfloat[]{\includegraphics[width=.42\textwidth]{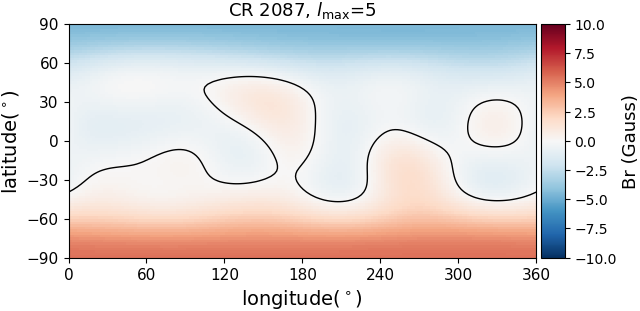}}
\caption{Synoptic GONG magnetograms during solar cycle maximum, CR 2159 (\textit{top left}) and cycle 
minimum, CR 2087 (\textit{top right}) 
followed by the spherical harmonic reconstructions with $l_\mathrm{max}$~=~150, 20, 10, 
and 5 respectively (\textit{second row to bottom}). The magnetic maps are saturated to {different values of} $B_\mathrm{r}$, 
to highlight the surface magnetic features.}
\label{fig1}
\end{figure*}

In 3D MHD solar wind models such as AWSoM, the input stellar surface magnetic field 
ensures that the model includes the correct magnetic topology of the stellar wind. 
In the case of the Sun, multiple solar observatories produce high-resolution synoptic magnetograms
which can be used as an input \citep{riley14}. 
Stellar wind models use {low-resolution} magnetic maps of stars as input
\citep{vidotto11,vidotto14,nicholson16,alvaradogomez16,alvaradogomez16b,garraffo16,dualta19},
which are reconstructed using the technique of Zeeman Doppler imaging
(ZDI; \citealt{semel89}). This imagining technique reconstructs the large-scale field 
using spectropolarimetric observations, where the field is typically described using spherical harmonic 
expansion. Alternatively, solar magnetograms are sometimes used as a proxy for a given Sun-like star and 
are scaled to its magnetic field and 
activity \citep{dong18}. One disadvantage of using {ZDI} stellar magnetic maps is their resolution. 
A typical stellar magnetic map is reconstructed up to a spherical harmonics degree, 
$l_\mathrm{max}$, of 5-10, while a solar magnetogram can have $l_\mathrm{max}\geq$100.
It is not known how the resolution of the magnetograms and the use of the Sun as a stellar proxy 
influence stellar wind properties determined from AWSoM simulations. 

In this study we validate AWSoM under low-resolution input conditions, which is an important {pre-requisite for the use of AWSoM in stellar cases}. 
We investigate whether or not AWSoM solar wind simulations under low-resolution input conditions 
can reproduce observed ACE \footnote{http://www.srl.caltech.edu/ACE/ASC/. Data accessed in 
October 2019.} solar wind properties at 1 AU. Under high-resolution input conditions, 
AWSoM wind properties show strong agreement with observed wind properties \citep{oran13,sachdeva19}. 
We carry out wind simulations using low-resolution input magnetograms 
and a varying $S_\mathrm{A}/B$ ratio to investigate the sensitivity of these two 
input parameters in determining wind properties.
Low-resolution magnetograms are obtained by performing spherical harmonic decompositions of high-resolution solar 
Global Oscillation Network Group (GONG) \footnote{https://gong.nso.edu/data/magmap/crmap.html} magnetograms 
for $l_\mathrm{max}$~=~150, 20, 10, and 5. We also obtain different values of
$S_\mathrm{A}/B$ from  far ultraviolet (FUV) spectral lines. The different values of $l_\mathrm{max}$ and $S_\mathrm{A}/B$ 
are used to create two grids of AWSoM wind simulations during minimum and maximum of the solar cycle. 
Additionally, we also use ZDI maps of three solar analogues as a replacement for the solar magnetic field
to investigate whether or not input magnetograms of stars with similar properties can be used as a proxy. 
In Section 2,  the wind model is introduced. In Section 3, we describe our grid of simulations. In Sections 4 and 5, 
we discuss our results and conclusions.
\section{Model description}
We use the data-driven AWSoM model of the 3D MHD code Block Adaptive Tree Solar Roe-Type Upwind Scheme
(BATS-R-US;\citealt{powell99}), which is publicly available under the Space Weather Modelling 
Framework (SWMF; \citealt{toth12}).
Alfv\'en wave partial reflection and dissipation lead to the heating of the plasma, thus no polytropic heating function is
required in this model. Thermal and magnetic pressure gradients lead to acceleration of the wind. 
The model incorporates two energy equations for protons and electrons with the same proton and electron 
velocities. In addition to radiative cooling, collisional heat conduction {\citep{spitzer56}}
is included near the star ($\leq$5 $R_\sun$) and 
collisionless  heat conduction {\citep{hollweg78}} is adopted far away from the star (>5$R_\sun$).  

The simulation framework consists of multiple modules. Here, we use the solar corona (SC) and the inner 
heliosphere (IH) module. The simulation setup for the SC module consists of a 3D spherical grid with an inner boundary immediately above
the stellar radius in the upper chromosphere (default at $\geq$1 $R_\odot$) and the outer boundary is at
a distance of 25 $R_\mathrm\sun$. {To resolve the transition region, the heat conduction and 
radiative cooling rates are artificially modified as discussed in detail by \citet{sokolov13}}.
The IH module starts at 18 $R_\mathrm\sun$ 
and extends beyond 1 AU. There is a coupling overlap between the two modules. 
{The simulation uses spherical block-adaptive grid in SC from 1 $R_\sun$ to 24 $R_\sun$ (grid blocks 
consist of 6$\times$4$\times$4 mesh cells) and Cartesian grid in IH (grid blocks consist of 4$\times$4$\times$4 mesh cells). 
The smallest cell size is ~0.001 $R_\sun$ near the star and ~1 $R_\sun$ at the SC outer boundary. 
In IH, the smallest cell is ~0.1 $R_\sun$ and largest cell is ~8 $R_\sun$. 
For both SC and IH, adaptive mesh refinement (AMR) is performed to resolve the current sheets in the simulation domain}
(for a detailed description of the model, see \citealt{sokolov13} and \citet{vanderholst14}).
\begin{table}
\centering
\caption{\label{inputparameters}Input parameters.}
\begin{tabular}{cc}
\hline
\hline
parameters&value\\
\hline
$S_\mathrm{A}/B$&1.1$\times$10$^6$ W~m$^{-2}$~T$^{-1}$\\
$\rho$&3$\times$10$^{-11}$kg~m$^{-3}$\\
T&50,000 K\\
$L_\bot\sqrt{B}$&1.5$\times$10$^5$~m~$\sqrt{\mathrm{{T}}}$\\
$h_\mathrm{S}$&0.17\\
\hline
\end{tabular}
\end{table} 

The solar or stellar surface magnetic field is one of the key lower boundary conditions. 
A potential field extrapolation is carried out using a potential field source surface (PFSS) model 
to obtain the initial magnetic field condition in the simulation domain. {The source surface
radius is kept at 2.5 $R_\sun$}.
The Alfv\'en wave Poynting flux is injected at the base of the simulation to heat and accelerate the wind.
{The Alfv\'en wave Poynting flux $S_\mathrm{A}/B$ is usually set to be 1.1$\times$10$^6$ W~m$^{-2}$~T$^{-1}$. Here,
we investigate how a change in the value of $S_\mathrm{A}/B$ and the resolution of the solar surface magnetic
field alter the simulated wind properties}.

The model includes multiple other input parameters, such as 
the {base density and temperature, the stochastic heating term, and the transverse correlation length of the
Alfv\'en wave. The base density and temperature are {fixed at} 
3$\times$10$^{-11}$kg~m$^{-3}$ and 50,000 K respectively. 
Many stellar wind models use the temperature at the lower boundary as a free parameter and scale this value to 
stars based on measurements of coronal temperatures, which have been observed to 
depend on the star's activity level \citep{holzwarth07,johnstone15c}. 
However, the base temperature in our model is not the coronal temperature, 
and our results are not strongly sensitive to the choice of this value.
The stochastic heating term $h_\mathrm{S}$ was taken to be 0.17 and determines the energy partitioning between 
the electrons and protons in the model, which is from a linear wave theory by \citet{chandran11}
and is kept constant in this work. 
The transverse correlation length of the Alfv\'en waves $L_\bot$ in the plane perpendicular to the magnetic
field $B$ is responsible for partial reflection of forward-propagating Alfv\'en waves required to form the turbulent cascade.
{The value of $L_\bot\sqrt{B}$ used in this model is 1.5$\times$10$^5$~m~$\sqrt{\mathrm{{T}}}$ and is an
adjustable input parameter}.  
}

We use two input magnetograms to simulate wind properties at solar cycle maximum and minimum, 
Carrington rotation CR 2159 and CR 2087, respectively. The magnetograms are input into the simulations in the form of
spherical harmonic decomposition. The maximum spherical harmonics degree considered
determines the resolution of the magnetogram and therefore the minimum size of the 
magnetic features on the stellar surface.
For the highest resolution simulation in this study, the spherical harmonics degree 
$l_\mathrm{max}$ is truncated to 150 and {$S_\mathrm{A}/B$=1.1$\times$10$^6$~W~m$^{-2}$~T$^{-1}$.}
 The {rest of the input parameters are} 
listed in Table \ref{inputparameters} and taken from \citet{vanderholst14}. 
%%%%
%%%%
%%%%
%%%%
\section{Two grids of low-resolution solar wind simulations}
To investigate the dependence of solar wind properties on low-resolution data, we created two grid of simulations. Only 
the input magnetogram resolution 
and the Alfv\'en wave Poynting flux ratio ($S_\mathrm{A}/B$) were altered and the rest of the input 
boundary conditions (Table \ref{inputparameters}) were kept constant. The two grids of simulations, the first 
grid during solar cycle maximum (CR 2159) and the second during minimum (CR 2087), were created by carrying out 
spherical harmonic decompositions of the input magnetogram for four different values of the maximum harmonics degree, 
$l_\mathrm{max}$~=~150, 20, 10, and 5. 
Additionally, four different values of the $S_\mathrm{A}/B$ ratio were used, where one $S_\mathrm{A}/B$ was taken from \citet{sokolov13} and the remaining three 
$S_\mathrm{A}/B$ values were determined from three different 
FUV spectral lines of the solar twin 18 Sco. 
The grid setup is identical for both 
solar maximum and minimum. Furthermore, we explored the use of a proxy 
magnetogram by including ZDI large-scale magnetic maps of three solar analogues instead of an input solar 
magnetic field to AWSoM.   
\subsection{Spherical harmonics decomposition of CR 2159 and CR 2087}
Stellar magnetograms reconstructed using ZDI have a much lower resolution compared to solar magnetograms. 
The majority of the stellar magnetograms have $l_\mathrm{max}\leq$10. We used spherical harmonics decomposition 
on two different solar magnetograms to bring their resolution down to ZDI level. The magnetograms were obtained during solar cycle maximum and minimum, 
CR 2159 and CR 2087 respectively (Fig. \ref{fig1}, \textit{top}). The synoptic magnetograms were obtained using GONG,
where the photospheric field is considered to be purely radial. We carried out spherical harmonic decompositions on the 
synoptic magnetograms using the PFSS model available in BATS-R-US \citep{toth2011}. The output is a set of complex spherical harmonics coefficients 
${\alpha_\mathrm{lm}}$ for a range of spherical harmonics degrees $l$= 0,1, ...., $l_\mathrm{max}$. 
 
The $\alpha_{lm}$ coefficients were used to calculate $B_{r}(\theta,\phi)$ for $l_{max}$= 150, 20, 10, and 5 
based on equation \ref{eq1} \citep{vidotto16},
\begin{equation}
B_r(\theta,\phi)=\sum_{l=1}^{l_\mathrm{max}}\sum_{m=0}^{l}\alpha_\mathrm{lm}Y_\mathrm{lm}(\theta,\phi),
\label{eq1}
\end{equation} 
\begin{equation}
Y_\mathrm{lm}= c_\mathrm{lm}P_\mathrm{lm}(\cos\theta)e^{im\phi},
\label{eq2}
\end{equation}
\begin{equation}
c_\mathrm{lm}= \sqrt{\frac{2l+1}{4\pi}\frac{(l+m)!}{(l-m)!}},
\label{eq3}
\end{equation}
where $P_\mathrm{lm}(\cos\theta)$ is the Legrende polynomial associated with degree $l$ and order $m$ and $c_\mathrm{lm}$ is a normalisation constant. 
The summation is carried out over 1$\leq l\leq l_\mathrm{max}$ and $-l\leq m \leq l$. The above equations are also implemented 
in the ZDI technique \citep{donati06}, where large-scale stellar surface magnetic geometry 
is reconstructed by solving for $B_{r}(\theta,\phi)$ \footnote{ZDI studies also reconstruct the meridional 
$B_{\phi}(\theta,\phi)$ and azimuthal field $B_{\theta}(\theta,\phi)$, which are not used in this work.}, 
often using lower values of spherical harmonics order, $l_\mathrm{max}\leq$10. 
We used Equations \ref{eq1}-\ref{eq3} to obtain 
low-resolution magnetograms by restricting $l_\mathrm{max}$ to 150, 20, 10, and 5. 

Figure \ref{fig1} shows the synoptic GONG magnetograms 
followed by the low-resolution reconstructions for both CR 2159 (left column) 
and CR 2087 (right column). 
The magnetograms reconstructed by restricting $l_\mathrm{max}$ to 5 and 10 are representative of 
solar large-scale magnetograms and can be considered similar to a ZDI magnetic map of the Sun \citep{fares17}.
The radial magnetic field geometry was extrapolated into a 3D coronal magnetic field by using a 
PFSS solution as a starting condition for the simulations. Either spherical harmonics or a 
finite difference potential field solver (FDIPS) can be used. \citet{toth2011} showed
that it is preferable to use FDIPS over spherical harmonics as ring patterns are sometimes seen 
near strong magnetic field regions when the spherical harmonics technique is used,
specifically for higher values of $l_\mathrm{max}$. We used spherical harmonics to be consistent 
with ZDI large-scale stellar magnetograms. Additionally, we are interested in low values of $l_\mathrm{max}$, where the 
impact is minimal.
%%%%%%%
%%%%%%%
\subsection{Alfv\'en wave Poynting flux to B ratio ($S_\mathrm{A}/B$)}
\begin{figure}
\centering
\includegraphics[width=.5\textwidth]{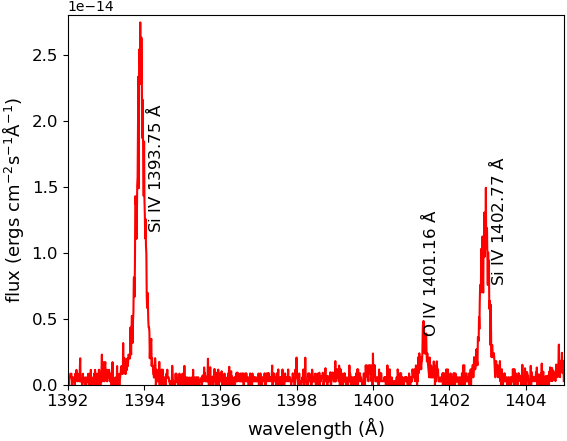}
\caption{HST FUV spectrum of 18 Sco. The spectral lines used in this analysis are marked.}
\label{fuv}
\end{figure}
The Poynting flux to B ratio ($S_\mathrm{A}/B$) is a key input parameter that characterises 
the heating and acceleration of the wind. For a solar wind
simulation using AWSoM, the $S_\mathrm{A}/B$ ratio was set by \citet{sokolov13} to be 1.1$\mathrm\times$ 10$^6$ 
W~m$^{-2}$~T$^{-1}$. In stellar wind modelling using AWSoM, $S_\mathrm{A}/B$ is sometimes 
adapted based on scaling laws between magnetic flux and X-ray flux \citep{garraffo16,dong18}. 
In this work, we investigated how the $S_\mathrm{A}/B$ ratio 
influences the mass and angular momentum loss rates, and other wind properties such as wind velocity, density, and ram pressure. 

{In the case of the Sun, the Alfv\'en wave Poynting flux $S_\mathrm{A}$ can be determined if we know 
the Alfv\'en speed $V_\mathrm{A}$ and the wave energy density $w$,    
\begin{equation}
S_\mathrm{A}= V_\mathrm{A}w
\label{vaeq}
,\end{equation}
\begin{equation}
V_\mathrm{A} = B/{\sqrt{{\mu_0}\rho}}
\label{saeq}
,\end{equation}
under the assumption of equipartition of kinetic and thermal energies of Alfv\'en waves, the wave energy density $w$ can be expressed as,
\begin{equation}
w = \rho \delta{v}^2
\label{weq}
,\end{equation}
resulting in the following $S_\mathrm{A}/B$ ratio,
\begin{equation} 
S_\mathrm{A}/B={\rho\delta{v}^2}/{\sqrt{{\mu_0}\rho}}, 
\label{sab}
\end{equation}
}where $\rho$ is the base density, $\delta{v}^2$ is the turbulent perturbation, and $\mu_0$ is the magnetic permeability of free space. The turbulent 
perturbation is related to the non-thermal turbulent velocity, $\xi^2=\frac{1}{2}<\delta{v^2}>$ \citep{banerjee98}. 
If we know the non-thermal velocity and base density for a given star, we can estimate the $S_\mathrm{A}/B$ ratio. 
Both of these quantities can be estimated using FUV spectra of stars using spectral lines 
that are formed in the upper chromosphere or transition region. 
Works by \citet{banerjee98}, \citet{pagano}, \citet{wood97}, 
and \citet{oran17} have 
shown that the non-thermal velocities 
can be determined from FUV spectral line broadening. However the $S_\mathrm{A}/B$ determined
from FUV spectra will strongly depend on the spectral line used and can vary significantly even for the same star. 
Non-thermal velocities in the Sun can vary in a range of 10-30~km~s$^{-1}$ \citep{depontieu07}, where the distribution
peaks at 15~km~s$^{-1}$. 

Here, we kept the base density of the solar wind constant (Table \ref{inputparameters}) and only changed the value of 
the non-thermal velocity in Equation \ref{sab}. The non-thermal velocity was modified based on the analysis of three 
different spectral lines: \ion{Si}{IV} at 1402 $\AA$, \ion{Si}{IV} 1393 $\AA$, and \ion{O}{IV} 1401 \AA. 
A Hubble Space Telescope (HST) spectrum of the solar twin 18 Sco (HD146233) was used as a solar proxy 
(Fig. \ref{fuv}), instead of the Interface Region Imaging Spectrograph (IRIS) solar observations
to ensure that the non-thermal velocities used in this work have 
similar uncertainty level as for other stars. 
IRIS is also not a full disk instrument although it produces full disk mosaic of the Sun
once per month~\footnote{https://iris.lmsal.com/mosaic.html}. 
The star 18 Sco was chosen as it is a well-known solar twin with a similar rotation
rate as the Sun \citep{porto97}. 

We determined the non-thermal velocity by carrying out a double Gaussian fit to our three selected 
spectral lines, where the full width at half maximum (FWHM) of the narrow component of the fit gives $\xi$. 
{The non-thermal velocity is assumed to be purely due to transverse Alfv\'en waves and can be used to
determine the turbulent velocity perturbation $\delta{v}^2$\citep{oran17}.}
Figure \ref{si} shows the \ion{Si} {IV} line at 1393.75 \AA~and the the double Gaussian fit to the line. 
Table \ref{nonthermal} lists the $\xi$ for each of the three spectral lines used.
According to \citet{wood97}, the narrow component of the line profile accounts for the non-thermal velocity while the
broad component could be attributed to micro-flaring, though \citet{ayres} showed that the origin of the
broad component is not entirely clear and could be due to 
chromospheric bright points \citep{peter06}. However, we note that in red giants the enhanced broadening 
near the wings is attributed to both radial and tangential turbulence produced by Alfv\'en waves 
\citep{carpenter97,robinson98,airapetian10}.

We used Equation \ref{non} to determine the non-thermal velocity 
from the measured FWHM \citep{banerjee98,oran17}, which is then used to determine $S_\mathrm{A}/B$.
The FWHM is given by,
\begin{equation}
\mathrm{FWHM} = \sqrt{4 \ln 2 \left({\frac{\lambda}{c}}^2\right)\left(\frac{2k_BT_i}{M_i}+{\xi}^2\right)},
\label{non}
\end{equation}
where FWHM is the full width half maximum of the narrow component of the double Gaussian fit, $\lambda$ is the rest wavelength 
of the spectral line in \AA, $c$ is the speed of light in km~s$^{-1}$, $k_\mathrm{B}$ is the 
Boltzmann constant, $M_\mathrm{i}$ is the atomic mass of the element,
and $T_\mathrm{i}$  is the formation temperature in K.  We fitted both single and double Gaussian line profiles and used a
$\chi^2$ test to determine the goodness of fit. The fit is always better when a double Gaussian profile is used.

\begin{table}
\centering
\caption{Formation temperature and non-thermal 
velocity for the three FUV spectral lines and the corresponding $S_\mathrm{A}/B$ ratios.}
\label{nonthermal}
\begin{tabular}{ccccc}
\hline
\hline
spectral line&wavelength&$T_\mathrm{i}$&$\xi$&$S_\mathrm{A}/B$\\
&\AA&K&km~s$^{-1}$&W~m$^{-2}$~T$^{-1}$\\
\hline
\ion{Si}{IV}&1393.75&60,000&29.6&2.2 $\times$10$^6$\\ 
\ion{Si}{IV}&1402.77&60,000&26.6&2.0 $\times$10$^6$\\
\ion{O}{IV}&1401.16&50,000&16.0& 1.2 $\times$10$^6$\\
\hline
\end{tabular}
\end{table}
The non-thermal velocity determined using the \ion{O}{IV} line is in good agreement with the peak solar non-thermal velocity of  
15~km~s$^{-1}$ \citep{depontieu07}. The estimated $\xi$ using the \ion{Si}{IV} lines 
are much higher. According to \citet{phillips08} the non-thermal velocity might depend on the height above the solar limb. 
The formation temperature of the
\ion{O}{IV} line is 50,000~K, which is also the base temperature of our simulation grid. This non-thermal velocity
results in a $S_\mathrm{A}/B$  that is very close to the well calibrated $S_\mathrm{A}/B$ of 
\citet{sokolov13}. The formation temperatures of the \ion{Si}{} lines are slightly higher and lead to a 
higher $\xi$ as listed in Table \ref{nonthermal}. Investigation of solar non-thermal velocity at different heights by
\citet{banerjee98} shows that the non-thermal velocity could be as high as 46~km~s$^{-1}$ and changes with height. This could be 
linked to the damping of Alfv\'en waves as they move from the chromosphere to the corona due to wave reflection 
and dissipation. A detailed discussion on these different line formations is however
beyond the scope of this work.
\begin{figure}
\centering
\includegraphics[width=.5\textwidth]{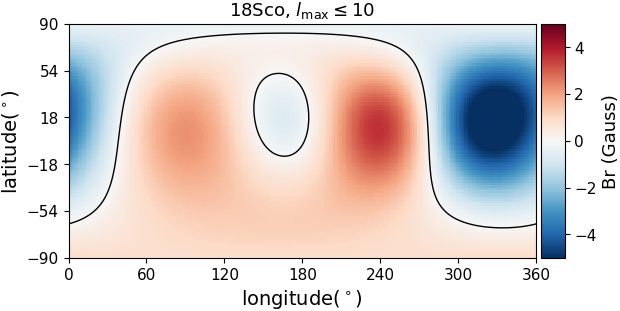}

\vspace{0.5cm}

\includegraphics[width=.5\textwidth]{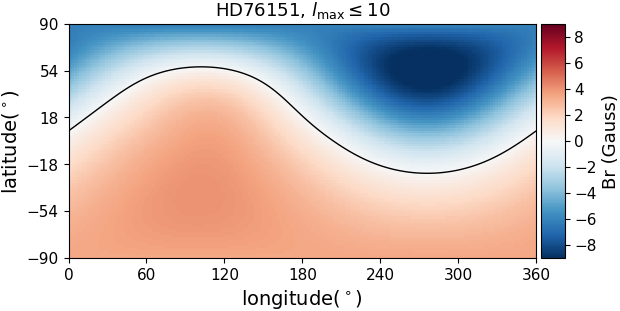}

\vspace{0.5cm}

\includegraphics[width=.5\textwidth]{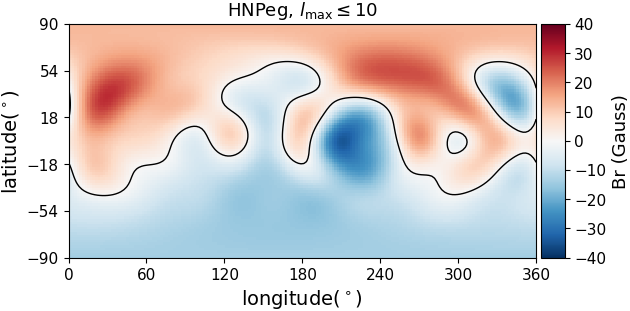}
\caption{ZDI large-scale magnetograms of 18 Sco, HD76151, and HN Peg (\textit{top} to \textit{bottom}).}
\label{hd76151}
\end{figure}

In the solar case, direct observations of the solar chromosphere and corona have lead to a detailed understanding of 
non-thermal velocities in the upper atmosphere. We can compare the non-thermal velocities obtained from FUV spectra with direct spatial observations. 
However, stellar observations lack the spatial and temporal resolution of the Sun. 
It becomes difficult to determine which out of the many non-thermal velocities available should be used to 
estimate $S_\mathrm{A}/B$. Therefore, we ran simulations of the solar wind 
by scaling $S_\mathrm{A}/B$  using the three different non-thermal velocities given 
in Table \ref{nonthermal} to investigate its influence on the wind properties.
%%%%%%%%%%%%%%%%%%%%%%%
\subsection{Stellar magnetic maps as a proxy for the solar magnetogram}
Currently, ZDI \citep{semel89,brown91,donati97,piskunov02,kochukhov02,folsom18} is the only technique that can 
reconstruct the surface magnetic geometries of stars. It is a tomographic technique that 
reconstructs the large-scale magnetic geometry of stars from 
circularly polarised spectropolarimetric observations. It is an inverse method where a magnetic map is reconstructed by 
inverting observed spectropolarimetric spectra, where the surface magnetic field is described as a combination of spherical 
harmonic components \citep{donati06} using the same equations as in Section 3.1 \citep[see][for more details]{folsom18}.

As ZDI only reconstructs the large-scale magnetic field, the magnetic maps are generally limited to $l_\mathrm{max}\leq10$. 
As a result, the small-scale magnetic field cancel out and the global magnetic field is 
much weaker than in typical solar magnetograms. These ZDI magnetic maps
are used as input magnetograms for stellar wind studies, where the global magnetic field strength 
is sometimes artificially increased to account for the
loss of small-scale features. Since ZDI magnetic maps are only available for a handful of stars,
it is often necessary to use a ZDI map from a star with similar parameters as a proxy, and scale it.
However, the magnetic geometry of any two stars is different. 
Furthermore, the magnetic geometry of active Sun-like stars can evolve over very short time-scales \citep{jeffers17,rosen16}. 
Even the solar large-scale magnetic geometry changes complexity over the 
solar cycle, although the complexity of the solar magnetic field does not
lead to any significant changes in the solar wind mass loss rate. It is not known if the same is true for 
very active Sun-like stars. 

Due to the availability of observational constraints, our Sun is the best test case to investigate whether or not ZDI magnetic maps of solar 
analogues can be used as a solar proxy. If the simulated solar wind properties cannot be reproduced using a ZDI map of a solar 
analogue as a solar proxy, then it is very unlikely that the use of the Sun as a proxy for a cool star such as an M dwarf 
is reliable. The three solar proxies used in this work are 18 Sco, HD 76151, and HN Peg. 
With a rotation period of 22.7 days, 18 Sco is the only solar twin for which a large-scale ZDI surface magnetic 
reconstruction is available \citep{petit08}. HD76151 is a solar mass star and 
is rotating slightly faster than the Sun with a rotation period of
20.5 days \citep{petit08}. HN Peg is a young solar analogue and is rotating much faster than the Sun at
4.6 days \citep{borosaikia15}. Table \ref{sample} lists the stellar parameters of the sample. The large-scale 
magnetic geometries of 18 Sco and HD76151 were reconstructed by \citet{petit08}. The spectropolarimetric 
data are available as part of the open-source archive POLARBASE \citep{petit14}. We applied 
ZDI \citep{folsom18} on the POLARBASE data to obtain the maps in Fig.~\ref{hd76151}. 
The magnetic map of HN Peg was taken from \citet{borosaikia15}. Figure~\ref{hd76151} shows the large-scale 
radial magnetic field of the sample, where each map was reconstructed with $l_\mathrm{max}\leq$10.  
\begin{table}
\centering
\caption{\label{sample} Stellar parameters of the sample. The masses and radii 
are taken from \citet{valenti05} and the rotation periods
are taken from \citet{petit08} and \citet{borosaikia15}.}
\begin{tabular}{cccccc}
\hline
\hline
name&mass&radius&inclination&P$_\mathrm{rot}$\\
&M$_\mathrm{\sun}$&R$_\mathrm{\sun}$&$^\circ$&days\\
\hline
18 Sco&0.98$\pm$0.13&1.022$\pm$0.018&70&22.7\\
HD76151&1.24$\pm$0.12&0.979$\pm$0.017&30&20.5\\
HN Peg&1.085$\pm$0.091&1.002$\pm$0.018&75&4.6\\
\hline
\end{tabular}
\end{table}
We used the magnetic maps in Fig. \ref{hd76151} instead of an input solar magnetogram and carried 
out steady-state wind simulations. The other input parameters, such as $S_\mathrm{A}/B$, density, and temperature,
were taken from Table \ref{inputparameters}.
\section{Results and Discussion}
\subsection{Properties of the solar wind during solar cycle minimum and maximum}
To determine the solar wind properties during cycle minimum and maximum, we carried out high-resolution steady state solar wind simulations
(CR 2087 and CR 2159) where the input boundary conditions (Table \ref{inputparameters}) 
and numerical setup are the same as in \citet{vanderholst14}. The only difference is the use of 
a magnetogram where $l_\mathrm{max}$ is restricted to 150 
for the input magnetic field map.
{Figure \ref{3dplot} shows the steady state solutions for the solar maximum and solar minimum cases.}
From the steady state solutions we determine the 
mass loss rate ($\dot{M}$), angular momentum loss rate ($\dot{J}$), 
wind velocity ($u_\mathrm{r}$), density ($\rho$), and ram pressure ($P_\mathrm{ram}$) at 1 AU. While the mass 
and angular momentum loss rates are discussed individually 
for solar cycle maximum and minimum, we combined the simulated cycle maximum and minimum 
data and explored the wind velocity, density, and ram pressure in terms of the fast and slow components of the wind.
\begin{figure*}
\centering
\includegraphics[width=.8\columnwidth]{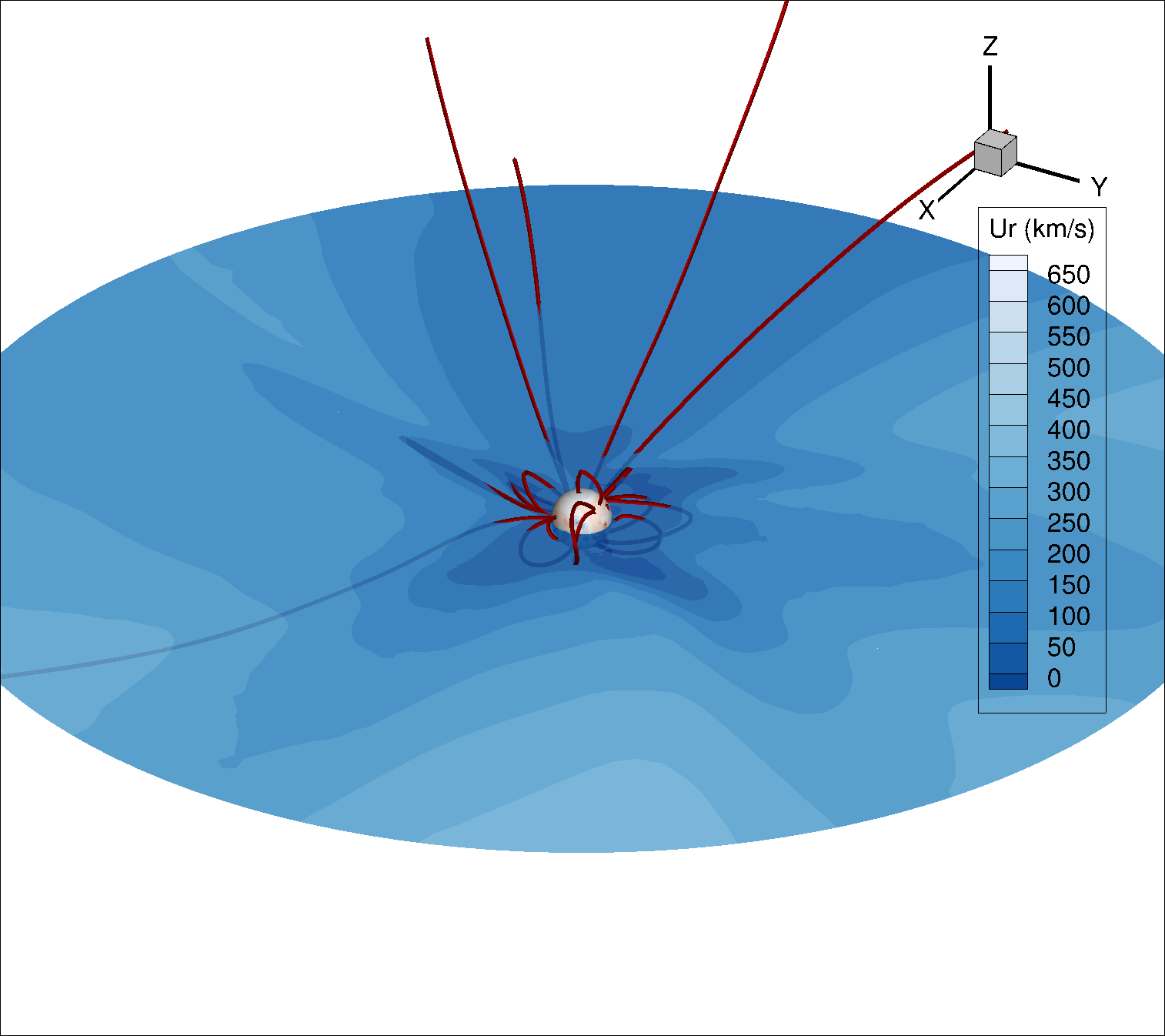}~~~\includegraphics[width=.8\columnwidth]{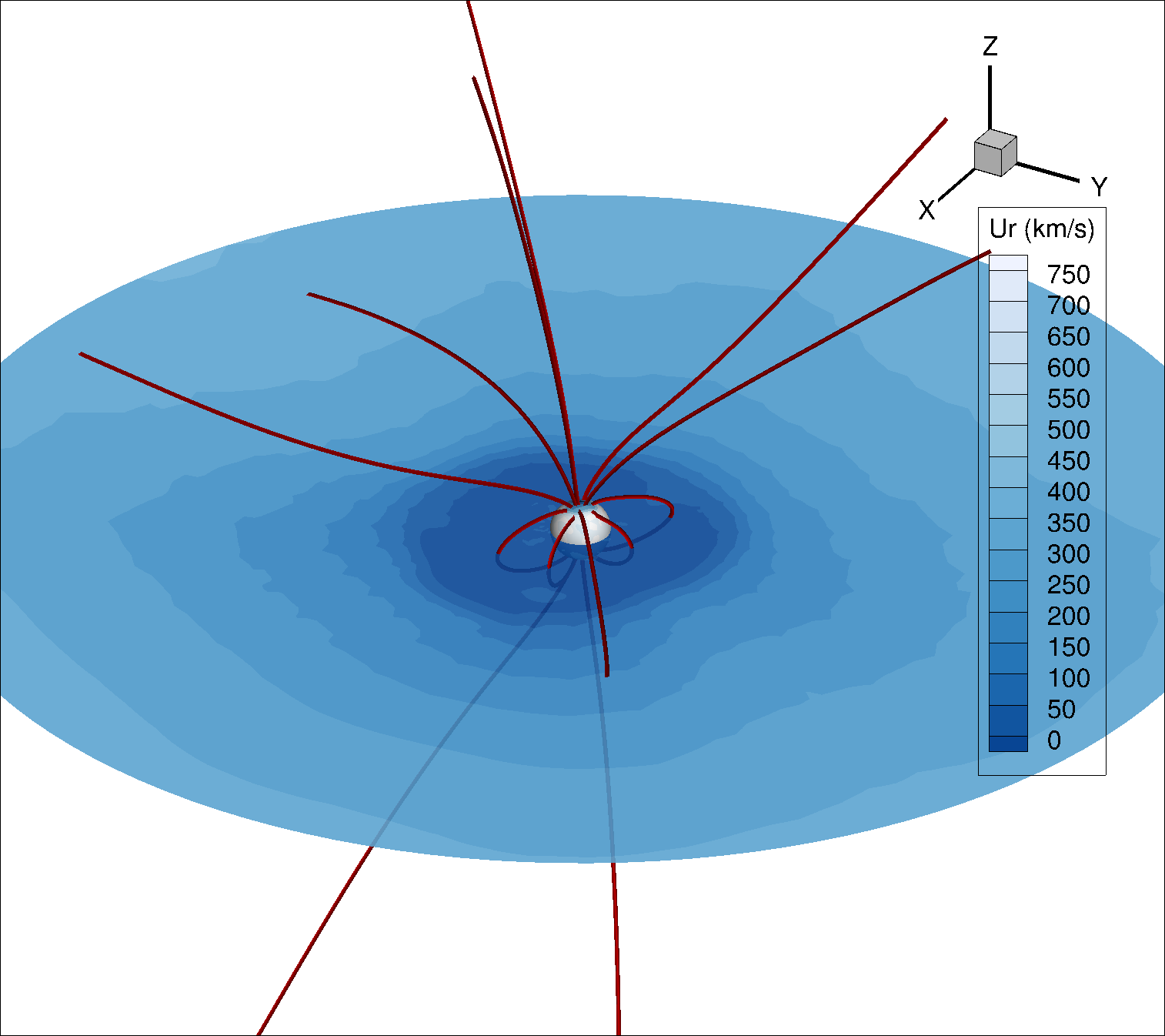}
\caption{Steady-state simulations for the solar maximum (CR 2159, left panel) and the solar minimum (CR 2087, right panel) cases. The  slice through 
\textit{z}=0 plane shows the radial velocity. Both open and closed magnetic field lines are shown in red streamlines. The surface magnetic field
geometry is shown on the solar surface as red and blue diverging contour.}
\label{3dplot}
\end{figure*}

The mass loss rate is determined by integrating the mass flux over a spherical surface, and is given by,
\begin{equation}
\dot M=\oint_S \rho u_\mathrm{r}\,dS
\label{mdoteq}
,\end{equation}
where $\rho$ is the density and $u_\mathrm{r}$ is the radial velocity of the wind at any given distance from the solar surface. 
The mass loss rate
of the wind is constant at any given distance from the Sun, except very close to the solar 
surface where not all magnetic field lines are open. 
The upper panel of Fig.~\ref{mdotjdot} shows the 
mass loss rate of the wind during solar cycle maximum and  minimum.
The global mass loss rate is 4.1$\times$10$^{-14}$~M$_\mathrm\sun$~yr$^{-1}$ during cycle maximum and
2.1$\times$10$^{-14}$~M$_\mathrm\sun$~yr$^{-1}$ during cycle minimum. Simulations of the solar wind during
solar cycle minimum and maximum by \citet{alvaradogomez16b} also agree with our
results, where these later authors spatially filtered the solar magnetograms to lower their resolution. 
{Low-resolution solar wind simulations were also carried out by \citet{dualta19} with mass loss rates~one magnitude weaker
than those obtained in this work. The use of different values in the input boundary conditions 
and different wind models could lead to such discrepancy.}
The mass loss rate of the Sun as observed by \textit{Ulysses} and \textit{Voyager} \citep{cohen11} shows 
a variability of a factor of approximately two, although it is not in phase with the minimum and maximum of the solar cycle. 
The mass loss rates obtained from our simulations fall within the observed variation.
The mass loss rates determined from our low-resolution simulations are discussed in the following section.
\begin{figure}
\centering
\includegraphics[width=.5\textwidth]{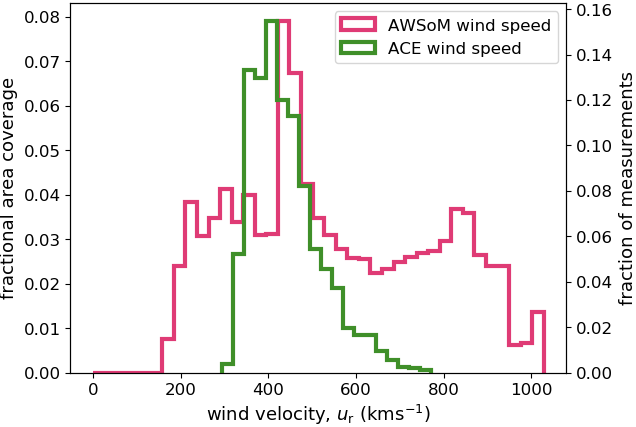}
\caption{Simulated and observed velocity of the wind, $u_\mathrm{r}$ in km~s$^{-1}$ at 1 AU. The combined 
$u_\mathrm{r}$ for both cycle maximum (CR 2159) and minimum (CR 2087) is shown. The ACE distribution consists of an entire 
year of data for CR 2159 and CR 2087. The left and right y axes show the 
factional area coverage of the AWSoM simulations and the fractional measurements of ACE respectively. 
}
\label{urplot}
\end{figure}
\begin{figure}
\centering
\includegraphics[width=.5\textwidth]{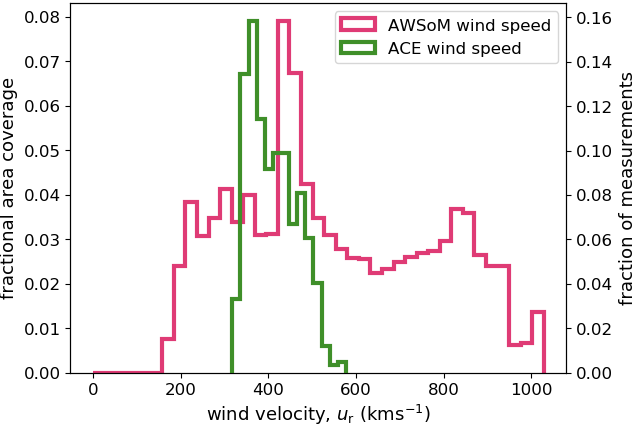}
\caption{Same as in Fig.~\ref{urplot} but for ACE data containing only CR 2159 and CR 2087.}
\label{urslowonly}
\end{figure}

Angular momentum is carried away from the star in two forms: the angular momentum held by the wind material and angular momentum
contained within the stressed magnetic field \citep{weber67}. The angular momentum loss rate is given by
\begin{equation}
\dot{J}= \oint_S\left[\frac{\varpi B_\mathrm{\phi}B_\mathrm{r}}{4\pi}+\varpi u_\mathrm{\phi}\rho u_\mathrm{r}\right]\,dS,
\label{jdoteq}
\end{equation}
where $\varpi=\sqrt{x^2+y^2}$ is the cylindrical radius, $\rho$ is the density, $B_\mathrm{r}$ and $B_\mathrm{\phi}$ are 
the magnetic field components, and $u_\mathrm{r}$ and $u_\mathrm{\phi}$ are the wind velocities. The subscripts 
$r$ and $\phi$ represent the radial and the azimuthal direction respectively. The first component of Equation 
\ref{jdoteq} is associated with the magnetic torque and the second component is the torque imparted by the plasma. 
As shown by \citet{vidotto14}, Equation \ref{jdoteq} is valid for stellar magnetic field geometries
that lack symmetry. The solar magnetic field is not always axisymmetric during the solar cycle {\citep{derosa12}}. 
Additionally, ZDI studies have shown that Sun-like stars often exhibit non-axisymmetric 
magnetic fields. For this reason, the well-known
relationship of angular momentum loss rates by \citet{weber67}, which is only 
applicable for axisymmetric systems, is not used here. 

During solar cycle maximum, the average AWSoM angular momentum loss is 4.0$\times$10$^{30}$~erg, while during cycle 
minimum it is 3.0$\times$10$^{30}$~erg (the lower panel of Fig.~\ref{mdotjdot} shows the angular 
momentum loss rate for the Sun during solar cycle maximum and minimum).
The angular momentum loss rates were obtained from the highest 
resolution  magnetogram used in this work, $l_\mathrm{max}$=150. It is therefore not 
surprising that the angular momentum loss rate
for both solar minimum and maximum is a factor of three or four higher than the angular momentum loss in 
\citet{alvaradogomez16b}, where the authors used low-resolution, 
spatially filtered magnetograms. Additionally, such small differences between the results in this work and 
\citet{alvaradogomez16b} could also occur due to the use of different synoptic Carrington maps. 
The angular momentum loss rates determined using AWSoM are in strong agreement with \textit{Helios} observations 
by \citet{pizzo83}, although \citet{finley18} suggested that the angular momentum loss rate in \citet{pizzo83} could be 
underestimated due to positioning of the satellite. Our values are also within the same magnitude as those determined
by \citet{finley18} using their open flux method.  
However our results are a magnitude higher than the 
angular momentum loss rates determined using 3D wind simulations of \citet{finley18a}, and \citet{reville17}, 
which \citet{finley18} 
attribute to their use of a polytropic equation of state instead of Alfv\'en wave heating.   
\begin{figure}
\centering
\includegraphics[width=.5\textwidth]{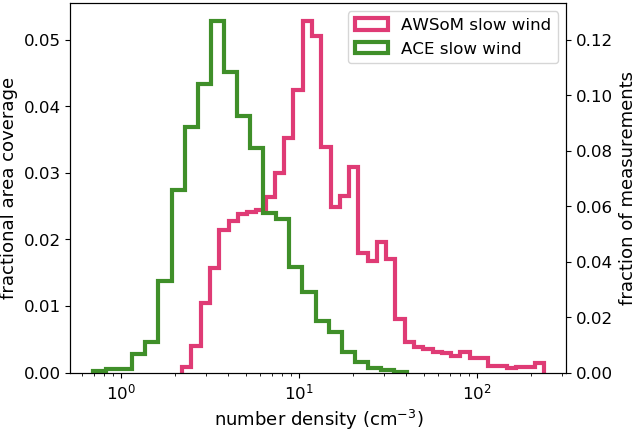}\\
\includegraphics[width=.5\textwidth]{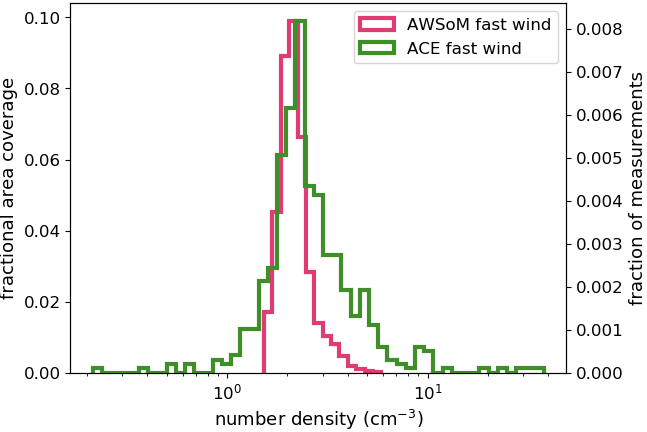}
\caption{Simulated and observed proton number densities for both slow (\textit{top}) and fast (\textit{bottom}) component of the wind at 1 AU.
The left y-axis represents fractional area coverage of our AWSoM simulations and the right y-axis represents the fraction of ACE measurements.}
\label{densityplot}
\end{figure}

We combined the wind output of the two steady-state simulations (solar maximum and minimum) to study wind properties 
such as velocity, proton density, and ram pressure as a function of the observed ACE wind properties.
Distribution of wind velocity $u_\mathrm{r}$ at a distance of 1 AU for combined 
solar cycle minimum (CR 2087) and 
maximum (CR 2159) are shown in Fig. \ref{urplot}. The 
IH component of the simulation grid was invoked to generate the wind properties at that distance. 
The distribution of the hourly averaged ACE solar wind speeds during the same 
years as CR 2159 and CR 2087 is also shown in Fig.~\ref{urplot}. 
The full two years containing CR 2159 and CR 2087 are 
combined to obtain the ACE distribution in Fig.~\ref{urplot}.
The fast wind cutoff is made at $u_\mathrm{r}$=600~km~s$^{-1}$ in this work. 
The simulated slow wind peak is in good agreement with the observed ACE slow wind peak. 
\begin{figure}
\centering
\includegraphics[width=.5\textwidth]{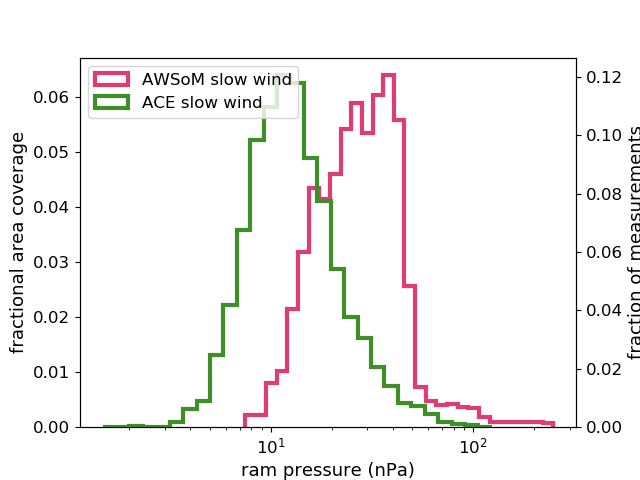}\\
\includegraphics[width=.5\textwidth]{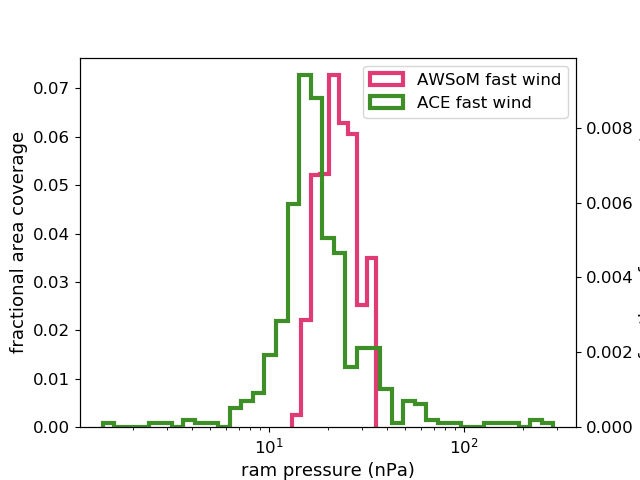}
\caption{Simulated and observed ram pressures due to the slow (\textit{top}) and fast (\textit{bottom}) components of the solar wind for the combined
cycle maximum and minimum simulations. The left y-axis shows the fractional area coverage of the AWSoM simulations and the right
y-axis shows the fraction of ACE measurements.}
\label{rampressure}
\end{figure}

Since the observations were taken by ACE, the distribution in Fig.~\ref{urplot} is biased towards the slow wind component. The 
ACE {satellite} is positioned at L1 in the equatorial plane and therefore mostly measures the slow component of the wind. 
\textit{Ulysses} measures both the slow and the fast component of the wind, but only limited measurements are available for the year that CR 2087 took place, during which time it was situated close to the 
equatorial plane. \textit{Ulysses} has no measurements from CR 2159. 
Multi-year observations taken by \textit{Ulysses} show that the fast wind speed is 
in good agreement with our results. The  median \textit{Ulysses} fast and slow wind speeds of 760~km~s$^{-1}$ and 400~km~s$^{-1}$
\citep{johnstone15a} are very similar to our median fast and slow wind speeds of 794~km~s$^{-1}$ and 391~km~s$^{-1}$ respectively.  
The median fast wind speed of ACE is 639~km~s$^{-1}$; however, this value could be biased because ACE does not 
have many observations of the fast wind. The median ACE wind speed is determined using all available data of the two
years containing CR 2159 and CR 2087. 
Figure \ref{urslowonly} shows the hourly averaged ACE wind velocities, {where only CR 2159 (January 2015) and CR 2087
(August 2009) data are included. The entire AWSoM distribution from Fig.~\ref{urplot} is also shown. 
During this period, no fast wind component was recorded by ACE. 
Therefore, we use all the data from the years that contain CR 2159 and CR 2087 (Fig. \ref{urplot}) to compare 
the model with observations of both slow and fast wind.}   

The proton density of the solar wind at a distance of 
1 AU for the combined solar cycle maximum (CR 2159) and minimum (CR 2087) simulations is shown in Fig. \ref{densityplot}. The density of the slow wind
is shown in the upper panel and the fast wind is shown in the lower panel.  The ACE proton density for the 
fast and slow wind is also shown in Fig \ref{densityplot}.
The proton density of the fast wind is lower than the proton density of the slow wind in our simulations, 
which is also seen in ACE observations. However very high slow wind proton densities are obtained in our simulations, which are not seen 
in the ACE data. The median proton density of the slow wind in our simulations is 12.7 cm$^{-3}$, which is about three times higher than 
the median ACE slow wind density (4.0 cm$^{-3}$). The agreement between AWSoM and ACE fast wind densities is 
better when compared to the slow wind. The median fast wind proton density in our simulations is 2.1 cm$^{-3}$ and the median ACE 
fast wind density is 2.3 cm$^{-3}$, although only very limited ACE fast wind measurements are available. 

We also calculated the ram pressure due to the solar wind as it is the dominant pressure component at a distance of 1 AU. 
The shape of planetary magnetospheres in the habitable zone of a Sun-like star strongly depends on the wind ram pressure.
The ram pressure due to the solar wind is calculated based on the following  equation, 
\begin{equation}
P_\mathrm{ram} = \rho u_\mathrm{r}^2,
\label{rameqn}
\end{equation}
where $\rho$ is the wind density in g~cm$^{-3}$ and $u_\mathrm{r}$ is the wind radial velocity in km~s$^{-1}$. 

Figure \ref{rampressure} shows the ram pressure $P_\mathrm{ram}$ distribution in nPa at a distance of 1 AU 
for both slow (Fig. \ref{rampressure}, upper panel) and fast components
(Fig. \ref{rampressure}, lower panel)  of the wind. The ram pressure calculated from ACE 
density and velocity measurements is also shown. There is no significant difference in the ram pressure for the slow and 
the fast wind. As density and wind speed have an inverse relation, they balance out in Equation \ref{rameqn}, resulting in 
similar contributions from both the fast and the slow wind components. 
The slight discrepancy in the ram pressure distribution between observations and simulations  
is most likely due to the high velocities ($\sim$ 1000~km~s$^{-1}$) of the fast wind component 
around the polar regions in the AWSoM simulations. No polar observations of the solar wind
exist to date, except for a few polar coronal hole measurements by \textit{Ulysses}. It is
therefore difficult to conclude how realistic the simulated polar wind speeds are. 
The AWSoM simulations lead to a median ram pressure of 31.8 nPa for the slow wind, which is 
higher than the ACE median ram pressure (12.4 nPa) by a factor of about 2.5. The median  AWSoM ram pressure for the fast wind in the simulations 
is 22.3 nPa, while the ACE observations lead to a median ram pressure of 17.2 nPa. 
\begin{figure*}
\centering
\includegraphics[width=.5\textwidth]{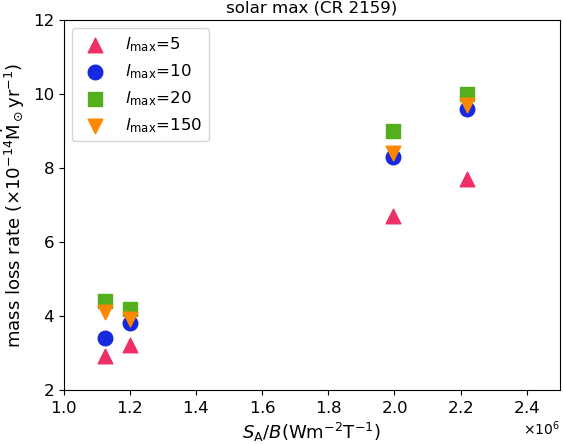}~~~\includegraphics[width=.5\textwidth]{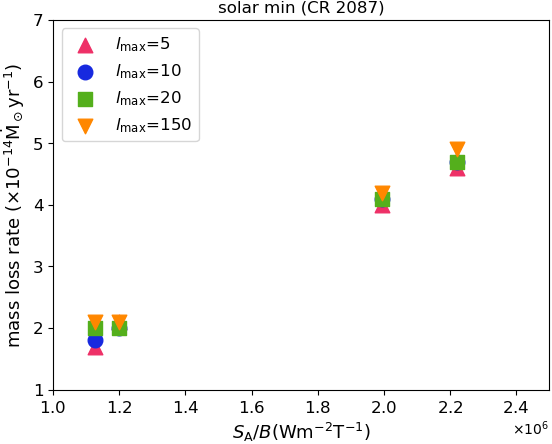}
\caption{Mass loss rate for the grid of simulations during cycle maximum (CR 2159) (\textit{left}) and minimum (CR 2087) (\textit{right}). 
The different colours represent different values of $l_\mathrm{max}$.}
\label{mdotgrid}
\end{figure*}

The discrepancies between simulated AWSoM and observed ACE wind velocities and 
proton densities could have several causes. 
It is well known in the solar community that although there is a general consensus between magnetograms from different solar observatories, 
there are still some discrepancies between their synoptic magnetic maps \citep{riley14}. 
Based on the choice of solar observatory for the input magnetogram, the final wind output could 
also vary \citep{gressl14}. Furthermore we cannot reliably observe the polar magnetic field and the polar 
field in the magnetograms is usually based on empirical models. This could also explain the very high wind 
velocities at the polar regions obtained in our simulations. 
Table \ref{prop} shows the median and mean solar wind properties during 
cycle minimum and maximum for the high-resolution solar wind 
simulations with $l_\mathrm{max}$=150 and $S_\mathrm{A}/B$=1.1~$\times$10$^6$~W~m$^{-2}$~T$^{-1}$. 
Caution should be taken regarding the fast wind 
properties of ACE as the satellite did not take enough observations of the fast component of the wind for the results to 
be statistically significant. 
\begin{table*}
\caption{Median and mean values of the wind speed, proton density, and ram pressure for the slow and the fast wind of both AWSoM simulations 
($l_\mathrm{max}$=150 and $S_\mathrm{A}/B$=1.1$\times$10$^6$~W~m$^{-2}$~T$^{-1}$) and ACE observations.}
\label{prop}
\centering
\begin{tabular}{ccccccccc}
\hline
\hline
&Median $u_\mathrm{r}$& Mean $u_\mathrm{r}$&Median $n_\mathrm{p}$&Mean $n_\mathrm{p}$&Median $P_\mathrm{ram}$&Mean $P_\mathrm{ram}$\\
&km~s$^{-1}$&km~s$^{-1}$&cm$^{-3}$&cm$^{-3}$&nPa&nPa\\
\hline
&&&slow wind&\\
AWSoM&391&380&12.7&20.3&31.8&35.3\\
ACE&420&430&4.0&5.1&12.4&15.0\\
\hline
&&&fast wind&\\
AWSoM&794&790&2.1&2.2&22.3&23.0\\
ACE&639&647&2.3&3.2&17.2&21.9\\
\hline
\end{tabular}
\end{table*}
%%%%%%%%%%%%
\subsection{Solar wind properties determined from our two grids of low-resolution simulations}
The two 4$\times$4 grids of simulations were created by altering the two key input parameters 
$l_\mathrm{max}$ and $S_\mathrm{A}/B$. The $S_\mathrm{A}/B$ value of 
1.1~$\times$10$^6$~W~m$^{-2}$~T$^{-1}$ is
the solar $S_\mathrm{A}/B$  taken from \citet{sokolov13}. The other three values of $S_\mathrm{A}/B$ were
determined from the HST spectra of 18 Sco. The other input parameters listed in Table \ref{inputparameters} were kept 
constant. The two grids represent solar cycle maximum (CR 2159) and minimum (CR 2087).

Figure \ref{mdotgrid} shows the mass loss rate 
for our two 4$\times$4 grids. During solar cycle maximum (left panel of Fig. \ref{mdotgrid}), 
the mass loss rate changes by a factor of $\leqslant$ 1.5 over a range of $l_\mathrm{max}$ for a given $S_\mathrm{A}/B$.
For example, keeping $S_\mathrm{A}/B$ constant at 1.1~$\times$10$^6$~W~m$^{-2}$~T$^{-1}$ 
and only changing the $l_\mathrm{max}$, the difference in the mass loss rate between the 
four simulations is a factor of about~1.5. However, if we keep $l_\mathrm{max}$ constant and use different values of 
$S_\mathrm{A}/B$, the mass loss rate can differ by a factor approximately 3. 
For the set of simulations where $l_\mathrm{max}$=5, the mass loss rate changes
by a factor of 2.6 over a range of $S_\mathrm{A}/B$. 
During solar cycle minimum (Fig. \ref{mdotgrid}, \textit{right}), the mass loss rate shows almost no variability for different values of $l_\mathrm{max}$
at a constant $S_\mathrm{A}/B$. The mass loss rate changes by a factor of about 2.7 or less for simulations with a constant $l_\mathrm{max}$ and varying $S_\mathrm{A}/B$.  
\begin{figure*}
\centering
\includegraphics[width=.5\textwidth]{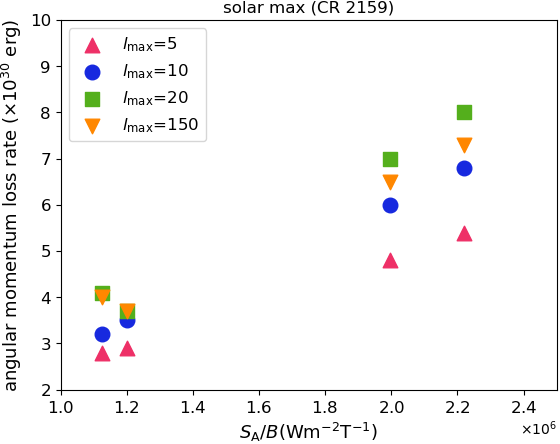}~~~\includegraphics[width=.5\textwidth]{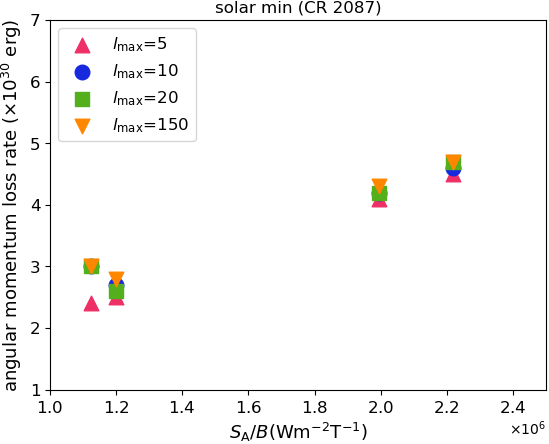}
\caption{Same as Fig. \ref{mdotgrid} except the angular momentum loss rate is shown instead of the mass loss rate.}
\label{jdotgrid}
\end{figure*}

Our results show that, depending on the activity level of the Sun,  
the resolution ($l_\mathrm{max}$) of the magnetic map has a small or negligible influence on the mass flux. During solar cycle maximum, the 
mass loss rate has a stronger dependence on the resolution (Table \ref{gridtable}) than 
during solar cycle minimum. Almost no variation is detected in the simulated mass loss rates
during solar cycle minimum (Table \ref{gridtable2}). The mass flux has a much stronger dependence on the $S_\mathrm{A}/B$ instead of $l_\mathrm{max}$.
Irrespective of the solar activity cycle, the mass loss rate changes by a factor of between two and three over a range of $S_\mathrm{A}/B$ at a given $l_\mathrm{max}$.

The angular momentum loss rate for the two grids during solar cycle maximum and minimum are shown in Fig. \ref{jdotgrid}. 
The variability in angular momentum loss for different values of $l_\mathrm{max}$ at a constant $S_\mathrm{A}/B$ during cycle maximum is 
a factor of $\leqslant$1.5. The variability increases to a factor of about  2 for different values of $S_\mathrm{A}/B$ at a constant 
$l_\mathrm{max}$ (Table \ref{gridtable}). During solar cycle minimum, the angular momentum shows negligible variations over a range
of $l_\mathrm{max}$ at a constant $S_\mathrm{A}/B$; it varies by a factor of $\leqslant$ 1.9 over a range of $S_\mathrm{A}/B$ at a constant
$l_\mathrm{max}$. The angular momentum loss shows similar dependence on $l_\mathrm{max}$ and $S_\mathrm{A}/B$ to the mass loss 
rate. Tables \ref{gridtable} and \ref{gridtable2} show the mass loss and the angular momentum loss rates 
for the two grids during cycle maximum and minimum respectively.

The mass loss and angular momentum loss rates show a slight decrease as the resolution lowers, as shown in Figs. \ref{mdotgrid} and \ref{jdotgrid}. 
This could be attributed to the loss of small-scale magnetic features for low values of $l_\mathrm{max}$, resulting in a
simpler field geometry. 
According to \citet{wangsheeley90}, the expansion of flux tubes from the photosphere to the corona
determines the wind density, temperature, velocity, and mass flux. The higher the expansion factor, the stronger the 
wind mass loss rate. The expansion factor increases for small-scale features which is another explanation for stronger
mass loss rates for higher values of $l_\mathrm{max}$. Furthermore, higher $l_\mathrm{max}$ also leads to stronger 
surface magnetic field, which leads to higher heating at the base leading to more mass loss. 
The higher expansion factor during solar cycle maximum, when the number of small-scale features is higher,
could also explain the increase in mass loss rate for CR 2159. 

{The impact of stellar magnetogram resolution was also investigated by \citet{jardine17}, who lowered the resolution of 
solar magnetograms using the same method as used in this work and used an empirical wind model to establish that 
the mass and angular momentum loss rates for a low-resolution magnetogram are within 5-20 $\%$ of the full resolution value. 
Since the large-scale dipole field is the key driver of mass and angular momentum loss in Sun-like stars, 
the resolution loss in ZDI does not have a significant influence on the 
mass or angular momentum loss rates \citep{reville15,see18}. 
However, the resolution of the magnetogram might have a stronger 
impact for slowly rotating stars with Rossby number $\>$2 \citep{see19}.}
 
Surprisingly, $l_\mathrm{max}$=20 leads to a marginally higher mass loss and angular momentum loss rate when compared to $l_\mathrm{max}$=150
during solar cycle maximum in CR 2159. As $l_\mathrm{max}$=150 has more closed small-scale magnetic 
regions it is expected to have the strongest mass and angular momentum loss. As shown in Equation~\ref{mdoteq}, $\dot{M}$ 
depends on the wind velocity $u_\mathrm{r}$ and  density $\rho$. As the solar wind moves outwards, 
the velocity increases and the number density decreases. 
Figure \ref{rhour} shows that during solar cycle maximum, the number density is slightly higher for $l_\mathrm{max}$=20 compared to 
$l_\mathrm{max}$=150. This could explain the slightly higher mass loss rate for $l_\mathrm{max}$=20.

The $S_\mathrm{A}/B$  has a stronger influence on the 
wind mass loss and stellar angular momentum loss compared to the choice of $l_\mathrm{max}$. 
This is not surprising since the Alfv\'en wave energy determines the heating and acceleration of the wind in the AWSoM model. 
This shows that robust determination of $S_\mathrm{A}/B$ is important for strong magnetic fields with complex field geometries. 
Our results also show that the \ion{O}{IV} line is a good tracer for $S_\mathrm{A}/B$ scaling. However, further 
investigations are needed to determine its suitability for other stars.

As the variation in mass loss and angular momentum loss is not significant over the given 
range of $l_\mathrm{max}$, we investigated how the wind speed, proton number density, and ram pressure are 
affected for the different values of $S_\mathrm{A}/B$ in Table \ref{nonthermal}.
We used the lowest ($l_\mathrm{max}$=5) and the highest ($l_\mathrm{max}$=150)  
resolution magnetograms for this purpose. These three wind properties for the fast and slow wind 
are determined from the combined solar maximum and minimum simulations.

Figure \ref{vgridmax} shows the wind speed for different values of $S_\mathrm{A}/B$ at $l_\mathrm{max}$=150 (upper panel) and 
$l_\mathrm{max}$=5 (lower panel), respectively. As expected, the distribution of the wind velocity is almost consistent
for the high-resolution ($l_\mathrm{max}$=150) and the low-resolution ($l_\mathrm{max}=5$) simulations. The wind velocity
shows a considerable variation with a varying $S_\mathrm{A}/B$. As the $S_\mathrm{A}/B$ increases the wind velocity of the fast
wind decreases, while the slow component does not show any considerable change in wind speeds.
Table \ref{prop1} and \ref{prop2} show the median and mean wind velocities at a distance of 1 AU for the distributions shown in
Fig. \ref{vgridmax}. 
\begin{figure*}
\centering
\includegraphics[width=.5\textwidth]{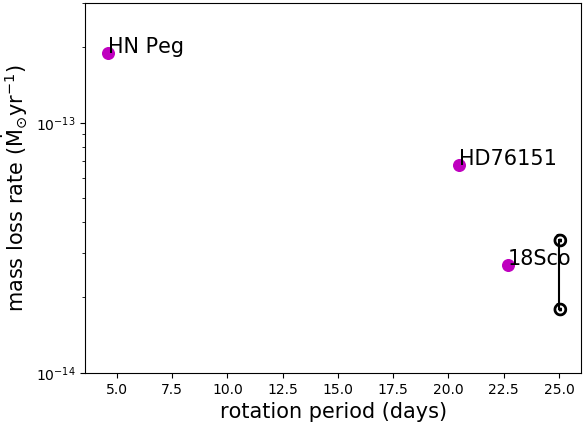}~~~\includegraphics[width=.5\textwidth]{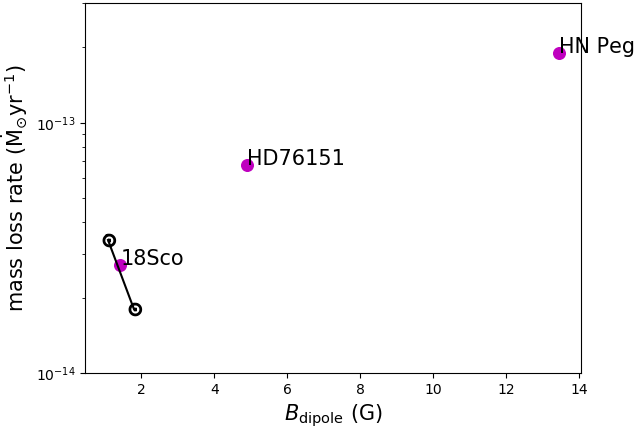}
\caption{Solar mass loss rate for different ZDI input magnetograms, shown in magenta, as a function of rotation period \textit{(left)} and 
dipolar field strength in Gauss \textit{(right)}. Solar symbols connected by the black vertical line shows the mass 
loss rate for the solar input magnetograms ($l_\mathrm{max}$=10) during solar cycle minimum and maximum. }
\label{zdimdot}
\end{figure*}
The proton density distribution of the slow and fast wind for $l_\mathrm{max}$=150 and 5 over a varying $S_\mathrm{A}/B$ 
is shown in Figs. \ref{rhogridslow} and \ref{rhogridfast} respectively. The proton density for both slow fast wind 
increases with increasing $S_\mathrm{A}/B$. The density increases by a factor of approximately two from 
$S_\mathrm{A}/B$=1.2 to 2.2. The median and mean values are shown in Table \ref{prop1} and \ref{prop2}.
The ram pressure shows a similar trend to the proton 
density distribution, as shown in Figs. \ref{ramgridslow} and \ref{ramgridfast}; it also shows a variation by a factor of approximately two depending on
the choice of $S_\mathrm{A}/B$. Although the mean and median wind velocity are not strongly influenced by the 
choice of $S_\mathrm{A}/B$, the density varies by a factor of about two, resulting in a corresponding change in ram pressure. 
The median and mean values of the ram pressure are tabulated in Tables \ref{prop1} and \ref{prop2}. 
Our results show that, similar to the mass loss and angular momentum loss rates, the density and ram pressure 
are more influenced by the $S_\mathrm{A}/B$ than by the resolution. Although the mean and median wind velocities are not strongly 
impacted by the variation of either $S_\mathrm{A}/B$ or $l_\mathrm{max}$, the distribution of the fast wind shows some
dependence on $S_\mathrm{A}/B$. For the solar case the variation is a factor of between two and three, but it could be much higher for a more active star.
%%%%%%%%%%%%%%%%%%%
%%%%%%%%%%%%%%%%%%%
%%%%%%%%%%%%%%%%%%%
\subsection{Solar wind properties determined using ZDI stellar magnetograms}
One problem often faced in stellar wind modelling is the lack of stellar magnetograms, 
as ZDI stellar magnetic maps are only available for $<$ 100 stars. To circumvent this problem,
solar magnetograms are sometimes used as a proxy for the stellar magnetic field in stellar wind modelling \citep{dong18}.
Unfortunately it is not known whether or not such approximations introduce any additional biases in the simulated 
wind properties. We investigated if the solar wind properties can be reproduced if we use ZDI 
magnetograms of Sun-like stars as the input for the solar magnetic field. This will allow us to have 
some insight into the usability of a magnetic map from one star (or even the sun) for a study of  the wind properties of another. 
We used large-scale ZDI magnetic maps of three solar analogues as a proxy for the solar magnetogram 
to carry out AWSoM solar wind simulations. The ZDI maps were used instead of the GONG magnetograms used in the previous 
section. Solar input boundary conditions are used, which are the 
same values as listed in Table \ref{inputparameters}. {Figure~\ref{3dplothnpeg} shows the velocity distribution of 
a steady-state simulation of one of the solar proxies HN Peg}.

Figure \ref{zdimdot} shows the mass loss of the solar wind for three different input ZDI magnetic maps reconstructed with
maximum spherical harmonic degree $l_\mathrm{max}$=10. The solar mass loss rate during cycle maximum (CR 2159) and minimum (CR 2087) 
for $l_\mathrm{max}$=10 and $S_\mathrm{A}/B$=1.1~$\times$10$^{6}$~W~m$^{-2}$~T$^{-1}$ is also shown. When a magnetic map of 
18 Sco is used as a solar proxy, the mass loss rate 
is in good agreement with the solar mass loss rate. The AWSoM solar wind simulation, where the solar magnetogram is
replaced by a large-scale magnetic map of the solar analogue HD76151, results in a mass loss rate 
that is more than three times the solar mass loss rate at cycle minimum. Finally, the AWSoM simulation, where a
large-scale magnetic map of the young solar analogue HN Peg is used as the input magnetogram, 
leads to a mass loss which is approximately ten times higher than the solar mass loss.

The angular momentum loss due to the solar wind, where these three ZDI large-scale magnetograms are used as input, is
shown in Fig. \ref{zdijdot}. The angular momentum loss rate of the Sun during cycle maximum and minimum where a magnetogram with 
$l_\mathrm{max}$=10 and $S_\mathrm{A}/B$=1.1~$\times$10$^{6}$~W~m$^{-2}$~T$^{-1}$ 
is used is also shown in Fig. \ref{zdijdot}. For both 18 Sco and HD76151 input
magnetic maps, the angular momentum loss is within one magnitude of the solar simulations in Fig. \ref{zdijdot}. However, there is 
a factor of approximately ten difference between the HN Peg simulations and the solar simulations in the previous section. 
Discrepancies in wind velocity, density,  and ram pressure between ACE observations and the three ZDI simulations are also
detected, as 
shown in Figs. \ref{urzdi}, \ref{rhozdi}, and \ref{ramzdi}.

Our results show that the large-scale magnetic map of the solar twin 18 Sco is a good solar proxy for wind simulations. 
The mass loss rate agrees strongly with both observed mass loss rates 
and our simulated mass loss rates for different values of $l_\mathrm{max}$.
The angular momentum loss is a factor of three lower than our simulations that use $l_\mathrm{max}$=10 
and $S_\mathrm{A}/B$=1.1~$\times$10$^{6}$~W~m$^{-2}$~T$^{-1}$, and also the
values obtained from \textit{Helios} observations \citep{pizzo83}. However, it falls within the range of angular momentum loss rates determined by \citet{finley18}
over the solar cycle. The dipole field strength of 18 Sco is similar to the solar dipolar field strength. The dipolar
component is primarily responsible for the wind mass loss and angular momentum loss \citep{reville15}.
HD76151 rotates at 20.5 days and when used in a solar wind simulation results in mass 
and angular momentum loss rates that are higher than those of 18 Sco. The mass loss rate 
agrees very well with cycle maximum solar wind simulations in the previous section 
where $S_\mathrm{A}/B$= 2.0~$\times$10$^6$~W~m$^{-2}$~T$^{-1}$. 
HD76151 has a stronger dipolar field when compared to the dipolar magnetic fields of both the Sun and 18 Sco. 
This could explain the mass and angular momentum loss rates 
being slightly higher than the solar magnetogram simulations for $l_\mathrm{max}$=10 and 
$S_\mathrm{max}/B$=1.1~$\times$10$^6$~W~m$^{-2}$~T$^{-1}$.
\begin{figure*}
\centering
\includegraphics[width=.5\textwidth]{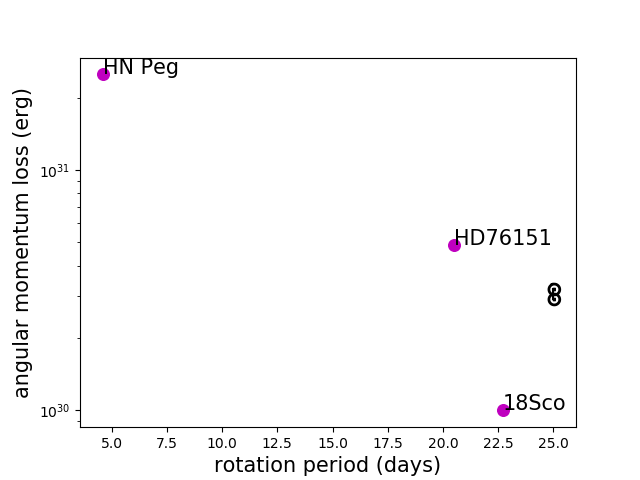}~~~\includegraphics[width=.5\textwidth]{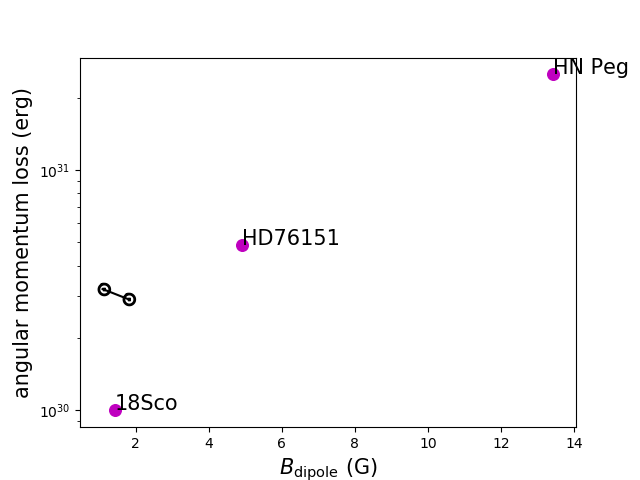}
\caption{Same as Fig. \ref{zdimdot}, except the angular momentum loss rate is shown instead of the mass loss rate.}
\label{zdijdot}
\end{figure*}

Unsurprisingly, the strongest mass and angular momentum loss rates are obtained for an input magnetogram of the young 
solar analogue HN Peg. The mass and angular momentum loss rates are a 
factor of approximately ten stronger than the solar cycle minimum mass and angular momentum loss rates 
(simulation carried out using $l_\mathrm{max}$=150, $S_\mathrm{A}/B$=1.1~$\times$10$^6$~W~m$^{-2}$~T$^{-1}$). 
This is much higher than the mass loss and angular momentum loss
rates obtained for our two 4$\times$4 grids of simulations. 
HN Peg is a young active star with an age of $\sim$250~Myr, which is known to harbour a strong toroidal magnetic
field \citep{borosaikia15}. Out of the multiple epochs of HN Peg data, we selected the epoch where the toroidal field is minimal (epoch 2009 in 
\citealt{borosaikia15}) with a strong poloidal field. The dipole field of HN Peg is much stronger than the solar
large-scale field and that of the other two proxies 18 Sco and HD76151 (Figs. \ref{zdimdot} and \ref{zdijdot}).
According to the \citet{finley18} the solar angular momentum loss varies by a factor of five 
over the sunspot cycle. The mass and angular momentum loss rates determined
from HN Peg magnetograms are still very high after taking the solar cycle variation into account. 

Our results show that a large-scale magnetogram of a given star can be used as an input to determine wind parameters for any solar-like star, provided 
the stellar parameters such as rotation and the dipolar field strength of the two stars are similar. It is not unusual to scale input magnetograms
based on the star's magnetic field strength. However, measuring dipolar field strengths of a star is not straightforward and can be subject to 
large uncertainties. In such cases, it is not clear if using scaling laws to scale the dipolar field is beneficial. 
Furthermore, errors in  parameters such as $S_\mathrm{A}/B$, density, and temperature could lead to added uncertainties 
(a detailed investigation with a bigger ZDI sample is beyond the scope of this work).
%%%%%%%%%%%%%%%%%%
%%%%%%CONCLUSION%%%%%%%%%%%
%%%%%%%%%%%%%%
\section{Conclusions}
We carried out solar wind simulations for two Carrington rotations CR 2159 and CR 2087, corresponding to 
solar cycle maximum and minimum, respectively, to investigate how the choice 
of solar input parameters influence the solar wind output. We 
lowered the resolution of solar magnetograms using spherical harmonic decomposition by varying the degree
$l_\mathrm{max}$=5, 10, 20, and 150. Additionally we altered the input Poynting flux to B ratio, $S_\mathrm{A}/B$, 
using non-thermal velocities determined from HST spectral lines. We used ACE wind properties at 1 AU
to validate our simulated solar wind properties during cycle maximum and minimum. 
Finally, we used stellar large-scale ZDI maps 
as proxies for the Sun to determine if the solar wind properties can be obtained using an input magnetogram of a solar 
analogue.   
 
Our key results can be summarised below:
\begin{itemize}
\item AWSoM solar wind simulations during solar cycle maximum (CR 2159) and minimum (CR 2087) reproduce solar wind properties
that agree with observed ACE wind properties to various extents. While the wind mass and angular momentum loss rates show good
agreement between wind simulations and observations, small discrepancies are detected in some other properties. The simulated 
wind velocities for the slow wind agree with the ACE slow wind velocities. Due to the lack of observations of the fast wind
it could not be established how well AWSoM reproduces the fast wind, specifically at polar regions, where the simulations
resulted in wind speeds of $\geq$1000~km~s$^{-1}$. However, the fast wind speeds obtained using AWSoM were validated against
\textit{Ulysses}, which puts confidence in the fast wind simulated in this work. 
The proton density and the ram pressure at 1 AU in the simulations 
is a factor of two to three higher than the ACE measurements. This slight discrepancy between the observations and the simulations 
could be due to a multitude of factors. The choice of solar observatory and the magnetogram itself could play a role.
Additionally, the polar magnetic field is not observed due to the Earth being at the ecliptic, which could lead to
discrepancies. Finally, the difference could be due to the model not accounting for heating mechanisms other than 
Alfv\'en-wave-driven heating. This shows that even for the solar case, we need more dedicated observations and modelling efforts.   

\item We investigated how the lack of robust high-resolution stellar data impacts the AWSoM wind properties. 
Our results show that $S_\mathrm{A}/B$
has a stronger influence on wind properties than the resolution ($l_\mathrm{max}$) of the input solar magnetogram. The 
resolution is more important during solar cycle maximum than cycle minimum. This shows that 
for a simpler less complex field the resolution does not 
matter as much as the $S_\mathrm{A}/B$ and large-scale ZDI magnetic maps can be used for stellar wind simulations. 
However, for stars with strong complex magnetic field geometries, resolution plays a small role but the contribution of 
$S_\mathrm{A}/B$ is still stronger. This shows that ZDI magnetograms provide reliable estimates on the underlying field and 
the limited resolution of ZDI is not the biggest concern. The choice of Alfv\'en energy is the dominant uncertainty. 

\item Finally, we also investigated whether the large-scale ZDI magnetic map of a solar analogue can be used as a proxy for the solar 
magnetogram. Due to the lack of stellar input magnetograms, it is assumed that the solar magnetogram can be used as a proxy for 
wind simulations of cool stars. Our results show that AWSoM can reproduce the solar wind properties using a ZDI magnetogram of
the solar twin 18 Sco instead of a solar magnetogram and using solar values for other input boundary conditions. However, 
the wind properties deviate when the magnetogram is replaced by rapidly rotating solar analogues. 
The wind properties vary by approximately one order of magnitude when the young solar analogue HN Peg is 
used as a proxy for the solar magnetogram. This shows that even for the same spectral type,
a moderate change in stellar parameters can lead to large uncertainties in the wind properties. These uncertainties could be even larger for
stars where the input boundary conditions are not as well constrained as for the Sun.
\end{itemize}
\begin{acknowledgements}
We thank the anonymous referee for their valuable comments and suggestions. 
S.B.S and T.L acknowledge funding via the Austrian Space Application Programme (ASAP) of the Austrian Research 
Promotion Agency (FFG) within ASAP11. S.B.S, T.L, C.P.J, K.G.K and M.G acknowledge the FWF NFN project S11601-N16 
and the sub-projects S11604-N16 and S11607-N16. V.S.A 
was supported by NASA Exobiology grant $\#$80NSSC17K0463,TESS Cycle 1 and by Sellers Exoplanetary Environments Collaboration 
(SEEC) Internal Scientist Funding Model (ISFM) at NASA GSFC.
\end{acknowledgements}
\bibliographystyle{aa}
\bibliography{ref}

\begin{thebibliography}{112}
\expandafter\ifx\csname natexlab\endcsname\relax\def\natexlab#1{#1}\fi

\bibitem[{{Airapetian} {et~al.}(2010){Airapetian}, {Carpenter}, \&
  {Ofman}}]{airapetian10}
{Airapetian}, V., {Carpenter}, K.~G., \& {Ofman}, L. 2010, \apj, 723, 1210

\bibitem[{{Airapetian} {et~al.}(2016){Airapetian}, {Glocer}, {Gronoff},
  {H{\'e}brard}, \& {Danchi}}]{Airapetian16b}
{Airapetian}, V.~S., {Glocer}, A., {Gronoff}, G., {H{\'e}brard}, E., \&
  {Danchi}, W. 2016, Nature Geoscience, 9, 452

\bibitem[{{Airapetian} {et~al.}(2017){Airapetian}, {Glocer}, {Khazanov},
  {Loyd}, {France}, {Sojka}, {Danchi}, \& {Liemohn}}]{airapetian17}
{Airapetian}, V.~S., {Glocer}, A., {Khazanov}, G.~V., {et~al.} 2017, \apjl,
  836, L3

\bibitem[{{Airapetian} \& {Usmanov}(2016)}]{airapetian16}
{Airapetian}, V.~S. \& {Usmanov}, A.~V. 2016, \apjl, 817, L24

\bibitem[{{Alazraki} \& {Couturier}(1971)}]{alazraki71}
{Alazraki}, G. \& {Couturier}, P. 1971, \aap, 13, 380

\bibitem[{{Alvarado-G{\'o}mez}
  {et~al.}(2016{\natexlab{a}}){Alvarado-G{\'o}mez}, {Hussain}, {Cohen},
  {Drake}, {Garraffo}, {Grunhut}, \& {Gombosi}}]{alvaradogomez16}
{Alvarado-G{\'o}mez}, J.~D., {Hussain}, G.~A.~J., {Cohen}, O., {et~al.}
  2016{\natexlab{a}}, \aap, 588, A28

\bibitem[{{Alvarado-G{\'o}mez}
  {et~al.}(2016{\natexlab{b}}){Alvarado-G{\'o}mez}, {Hussain}, {Cohen},
  {Drake}, {Garraffo}, {Grunhut}, \& {Gombosi}}]{alvaradogomez16b}
{Alvarado-G{\'o}mez}, J.~D., {Hussain}, G.~A.~J., {Cohen}, O., {et~al.}
  2016{\natexlab{b}}, \aap, 594, A95

\bibitem[{{Ayres}(2015)}]{ayres}
{Ayres}, T.~R. 2015, \aj, 149, 58

\bibitem[{{Banerjee} {et~al.}(1998){Banerjee}, {Teriaca}, {Doyle}, \&
  {Wilhelm}}]{banerjee98}
{Banerjee}, D., {Teriaca}, L., {Doyle}, J.~G., \& {Wilhelm}, K. 1998, \aap,
  339, 208

\bibitem[{{Barabash} {et~al.}(2007){Barabash}, {Fedorov}, {Sauvaud}, {Lundin},
  {Russell}, {Futaana}, {Zhang}, {Andersson}, {Brinkfeldt}, {Grigoriev},
  {Holmstr{\"o}m}, {Yamauchi}, {Asamura}, {Baumjohann}, {Lammer}, {Coates},
  {Kataria}, {Linder}, {Curtis}, {Hsieh}, {Sandel}, {Grande}, {Gunell},
  {Koskinen}, {Kallio}, {Riihel{\"a}}, {S{\"a}les}, {Schmidt}, {Kozyra},
  {Krupp}, {Fr{\"a}nz}, {Woch}, {Luhmann}, {McKenna-Lawlor}, {Mazelle},
  {Thocaven}, {Orsini}, {Cerulli-Irelli}, {Mura}, {Milillo}, {Maggi}, {Roelof},
  {Brandt}, {Szego}, {Winningham}, {Frahm}, {Scherrer}, {Sharber}, {Wurz}, \&
  {Bochsler}}]{Barabash07}
{Barabash}, S., {Fedorov}, A., {Sauvaud}, J.~J., {et~al.} 2007, \nat, 450, 650

\bibitem[{{Belcher} \& {Davis}(1971)}]{belcher71}
{Belcher}, J.~W. \& {Davis}, Jr., L. 1971, \jgr, 76, 3534

\bibitem[{{Blackman} \& {Tarduno}(2018)}]{blackman18}
{Blackman}, E.~G. \& {Tarduno}, J.~A. 2018, \mnras, 481, 5146

\bibitem[{{Boro Saikia} {et~al.}(2015){Boro Saikia}, {Jeffers}, {Petit},
  {Marsden}, {Morin}, \& {Folsom}}]{borosaikia15}
{Boro Saikia}, S., {Jeffers}, S.~V., {Petit}, P., {et~al.} 2015, \aap, 573, A17

\bibitem[{{Brown} {et~al.}(1991){Brown}, {Donati}, {Rees}, \&
  {Semel}}]{brown91}
{Brown}, S.~F., {Donati}, J.-F., {Rees}, D.~E., \& {Semel}, M. 1991, \aap, 250,
  463

\bibitem[{{Carpenter} \& {Robinson}(1997)}]{carpenter97}
{Carpenter}, K.~G. \& {Robinson}, R.~D. 1997, \apj, 479, 970

\bibitem[{{Chandran} {et~al.}(2011){Chandran}, {Dennis}, {Quataert}, \&
  {Bale}}]{chandran11}
{Chandran}, B. D.~G., {Dennis}, T.~J., {Quataert}, E., \& {Bale}, S.~D. 2011,
  \apj, 743, 197

\bibitem[{{Cohen}(2011)}]{cohen11}
{Cohen}, O. 2011, \mnras, 417, 2592

\bibitem[{{Cohen} {et~al.}(2007){Cohen}, {Sokolov}, {Roussev}, {Arge},
  {Manchester}, {Gombosi}, {Frazin}, {Park}, {Butala}, {Kamalabadi}, \&
  {Velli}}]{cohen07}
{Cohen}, O., {Sokolov}, I.~V., {Roussev}, I.~I., {et~al.} 2007, \apjl, 654,
  L163

\bibitem[{{Cranmer}(2009)}]{cranmer09}
{Cranmer}, S.~R. 2009, Living Reviews in Solar Physics, 6, 3

\bibitem[{{Cranmer} \& {Saar}(2011)}]{cranmer11}
{Cranmer}, S.~R. \& {Saar}, S.~H. 2011, \apj, 741, 54

\bibitem[{{Cranmer} {et~al.}(2007){Cranmer}, {van Ballegooijen}, \&
  {Edgar}}]{cranmer07}
{Cranmer}, S.~R., {van Ballegooijen}, A.~A., \& {Edgar}, R.~J. 2007, \apjs,
  171, 520

\bibitem[{{Cranmer} \& {Winebarger}(2019)}]{cranmer19}
{Cranmer}, S.~R. \& {Winebarger}, A.~R. 2019, \araa, 57, 157

\bibitem[{{De Pontieu} {et~al.}(2007){De Pontieu}, {McIntosh}, {Carlsson},
  {Hansteen}, {Tarbell}, {Schrijver}, {Title}, {Shine}, {Tsuneta}, {Katsukawa},
  {Ichimoto}, {Suematsu}, {Shimizu}, \& {Nagata}}]{depontieu07}
{De Pontieu}, B., {McIntosh}, S.~W., {Carlsson}, M., {et~al.} 2007, Science,
  318, 1574

\bibitem[{{DeRosa} {et~al.}(2012){DeRosa}, {Brun}, \& {Hoeksema}}]{derosa12}
{DeRosa}, M.~L., {Brun}, A.~S., \& {Hoeksema}, J.~T. 2012, \apj, 757, 96

\bibitem[{{Donati} \& {Brown}(1997)}]{donati97}
{Donati}, J.-F. \& {Brown}, S.~F. 1997, \aap, 326, 1135

\bibitem[{{Donati} {et~al.}(2006){Donati}, {Howarth}, {Jardine}, {Petit},
  {Catala}, {Landstreet}, {Bouret}, {Alecian}, {Barnes}, {Forveille},
  {Paletou}, \& {Manset}}]{donati06}
{Donati}, J.-F., {Howarth}, I.~D., {Jardine}, M.~M., {et~al.} 2006, \mnras,
  370, 629

\bibitem[{{Dong} {et~al.}(2018){Dong}, {Jin}, {Lingam}, {Airapetian}, {Ma}, \&
  {van der Holst}}]{dong18}
{Dong}, C., {Jin}, M., {Lingam}, M., {et~al.} 2018, Proceedings of the National
  Academy of Science, 115, 260

\bibitem[{{Drake} {et~al.}(1993){Drake}, {Simon}, \& {Brown}}]{drake93}
{Drake}, S.~A., {Simon}, T., \& {Brown}, A. 1993, \apj, 406, 247

\bibitem[{{Fichtinger} {et~al.}(2017){Fichtinger}, {G{\"u}del}, {Mutel},
  {Hallinan}, {Gaidos}, {Skinner}, {Lynch}, \& {Gayley}}]{fichtinger17}
{Fichtinger}, B., {G{\"u}del}, M., {Mutel}, R.~L., {et~al.} 2017, \aap, 599,
  A127

\bibitem[{{Finley} \& {Matt}(2018)}]{finley18a}
{Finley}, A.~J. \& {Matt}, S.~P. 2018, \apj, 854, 78

\bibitem[{{Finley} {et~al.}(2018){Finley}, {Matt}, \& {See}}]{finley18}
{Finley}, A.~J., {Matt}, S.~P., \& {See}, V. 2018, \apj, 864, 125

\bibitem[{{Folsom} {et~al.}(2018){Folsom}, {Bouvier}, {Petit}, {L{\`e}bre},
  {Amard}, {Palacios}, {Morin}, {Donati}, \& {Vidotto}}]{folsom18}
{Folsom}, C.~P., {Bouvier}, J., {Petit}, P., {et~al.} 2018, \mnras, 474, 4956

\bibitem[{{Gaidos} {et~al.}(2000){Gaidos}, {G{\"u}del}, \& {Blake}}]{gaidos00}
{Gaidos}, E.~J., {G{\"u}del}, M., \& {Blake}, G.~A. 2000, \grl, 27, 501

\bibitem[{{Garraffo} {et~al.}(2016){Garraffo}, {Drake}, \&
  {Cohen}}]{garraffo16}
{Garraffo}, C., {Drake}, J.~J., \& {Cohen}, O. 2016, \apjl, 833, L4

\bibitem[{{Gressl} {et~al.}(2014){Gressl}, {Veronig}, {Temmer}, {Odstr{\v
  c}il}, {Linker}, {Miki{\'c}}, \& {Riley}}]{gressl14}
{Gressl}, C., {Veronig}, A.~M., {Temmer}, M., {et~al.} 2014, \solphys, 289,
  1783

\bibitem[{{Hollweg}(1978)}]{hollweg78}
{Hollweg}, J.~V. 1978, Reviews of Geophysics and Space Physics, 16, 689

\bibitem[{{Holzwarth} \& {Jardine}(2007)}]{holzwarth07}
{Holzwarth}, V. \& {Jardine}, M. 2007, \aap, 463, 11

\bibitem[{{Jardine} {et~al.}(2017){Jardine}, {Vidotto}, \& {See}}]{jardine17}
{Jardine}, M., {Vidotto}, A.~A., \& {See}, V. 2017, \mnras, 465, L25

\bibitem[{{Jeffers} {et~al.}(2017){Jeffers}, {Boro Saikia}, {Barnes}, {Petit},
  {Marsden}, {Jardine}, {Vidotto}, \& {BCool Collaboration}}]{jeffers17}
{Jeffers}, S.~V., {Boro Saikia}, S., {Barnes}, J.~R., {et~al.} 2017, \mnras,
  471, L96

\bibitem[{{Johnstone}(2017)}]{johnstone17}
{Johnstone}, C.~P. 2017, \aap, 598, A24

\bibitem[{{Johnstone} \& {G{\"u}del}(2015)}]{johnstone15c}
{Johnstone}, C.~P. \& {G{\"u}del}, M. 2015, \aap, 578, A129

\bibitem[{{Johnstone} {et~al.}(2015{\natexlab{a}}){Johnstone}, {G{\"u}del},
  {Brott}, \& {L{\"u}ftinger}}]{johnstone15b}
{Johnstone}, C.~P., {G{\"u}del}, M., {Brott}, I., \& {L{\"u}ftinger}, T.
  2015{\natexlab{a}}, \aap, 577, A28

\bibitem[{{Johnstone} {et~al.}(2018){Johnstone}, {G{\"u}del}, {Lammer}, \&
  {Kislyakova}}]{Johnstone18}
{Johnstone}, C.~P., {G{\"u}del}, M., {Lammer}, H., \& {Kislyakova}, K.~G. 2018,
  \aap, 617, A107

\bibitem[{{Johnstone} {et~al.}(2015{\natexlab{b}}){Johnstone}, {G{\"u}del},
  {L{\"u}ftinger}, {Toth}, \& {Brott}}]{johnstone15a}
{Johnstone}, C.~P., {G{\"u}del}, M., {L{\"u}ftinger}, T., {Toth}, G., \&
  {Brott}, I. 2015{\natexlab{b}}, \aap, 577, A27

\bibitem[{{Kislyakova} {et~al.}(2014{\natexlab{a}}){Kislyakova},
  {Holmstr{\"o}m}, {Lammer}, {Odert}, \& {Khodachenko}}]{kislyakova14b}
{Kislyakova}, K.~G., {Holmstr{\"o}m}, M., {Lammer}, H., {Odert}, P., \&
  {Khodachenko}, M.~L. 2014{\natexlab{a}}, Science, 346, 981

\bibitem[{{Kislyakova} {et~al.}(2014{\natexlab{b}}){Kislyakova}, {Johnstone},
  {Odert}, {Erkaev}, {Lammer}, {L{\"u}ftinger}, {Holmstr{\"o}m}, {Khodachenko},
  \& {G{\"u}del}}]{kislyakova14}
{Kislyakova}, K.~G., {Johnstone}, C.~P., {Odert}, P., {et~al.}
  2014{\natexlab{b}}, \aap, 562, A116

\bibitem[{{Kochukhov} {et~al.}(2017){Kochukhov}, {Petit}, {Strassmeier},
  {Carroll}, {Fares}, {Folsom}, {Jeffers}, {Korhonen}, {Monnier}, {Morin},
  {Ros{\'e}n}, {Roettenbacher}, \& {Shulyak}}]{fares17}
{Kochukhov}, O., {Petit}, P., {Strassmeier}, K.~G., {et~al.} 2017,
  Astronomische Nachrichten, 338, 428

\bibitem[{{Kochukhov} \& {Piskunov}(2002)}]{kochukhov02}
{Kochukhov}, O. \& {Piskunov}, N. 2002, \aap, 388, 868

\bibitem[{{Kosugi} {et~al.}(2007){Kosugi}, {Matsuzaki}, {Sakao}, {Shimizu},
  {Sone}, {Tachikawa}, {Hashimoto}, {Minesugi}, {Ohnishi}, {Yamada}, {Tsuneta},
  {Hara}, {Ichimoto}, {Suematsu}, {Shimojo}, {Watanabe}, {Shimada}, {Davis},
  {Hill}, {Owens}, {Title}, {Culhane}, {Harra}, {Doschek}, \&
  {Golub}}]{kosugi07}
{Kosugi}, T., {Matsuzaki}, K., {Sakao}, T., {et~al.} 2007, \solphys, 243, 3

\bibitem[{{Krieger} {et~al.}(1973){Krieger}, {Timothy}, \&
  {Roelof}}]{krieger73}
{Krieger}, A.~S., {Timothy}, A.~F., \& {Roelof}, E.~C. 1973, \solphys, 29, 505

\bibitem[{{Lichtenegger} {et~al.}(2010){Lichtenegger}, {Lammer},
  {Grie{\ss}meier}, {Kulikov}, {von Paris}, {Hausleitner}, {Krauss}, \&
  {Rauer}}]{Lichtenegger10}
{Lichtenegger}, H.~I.~M., {Lammer}, H., {Grie{\ss}meier}, J.-M., {et~al.} 2010,
  \icarus, 210, 1

\bibitem[{{Lundin}(2011)}]{Lundin11}
{Lundin}, R. 2011, \ssr, 162, 309

\bibitem[{{Matsumoto} \& {Suzuki}(2012)}]{matsumoto12}
{Matsumoto}, T. \& {Suzuki}, T.~K. 2012, \apj, 749, 8

\bibitem[{{Matt} {et~al.}(2015){Matt}, {Brun}, {Baraffe}, {Bouvier}, \&
  {Chabrier}}]{matt15}
{Matt}, S.~P., {Brun}, A.~S., {Baraffe}, I., {Bouvier}, J., \& {Chabrier}, G.
  2015, \apjl, 799, L23

\bibitem[{{Matthaeus} {et~al.}(1999){Matthaeus}, {Zank}, {Oughton}, {Mullan},
  \& {Dmitruk}}]{matthaeus99}
{Matthaeus}, W.~H., {Zank}, G.~P., {Oughton}, S., {Mullan}, D.~J., \&
  {Dmitruk}, P. 1999, \apjl, 523, L93

\bibitem[{{McComas} {et~al.}(1998){McComas}, {Bame}, {Barker}, {Feldman},
  {Phillips}, {Riley}, \& {Griffee}}]{mccomas98}
{McComas}, D.~J., {Bame}, S.~J., {Barker}, P., {et~al.} 1998, \ssr, 86, 563

\bibitem[{{McComas} {et~al.}(2003){McComas}, {Elliott}, {Schwadron}, {Gosling},
  {Skoug}, \& {Goldstein}}]{mccomas03}
{McComas}, D.~J., {Elliott}, H.~A., {Schwadron}, N.~A., {et~al.} 2003, \grl,
  30, 1517

\bibitem[{{McIntosh} {et~al.}(2011){McIntosh}, {de Pontieu}, {Carlsson},
  {Hansteen}, {Boerner}, \& {Goossens}}]{mcintosh11}
{McIntosh}, S.~W., {de Pontieu}, B., {Carlsson}, M., {et~al.} 2011, \nat, 475,
  477

\bibitem[{{Nicholson} {et~al.}(2016){Nicholson}, {Vidotto}, {Mengel},
  {Brookshaw}, {Carter}, {Petit}, {Marsden}, {Jeffers}, {Fares}, \& {BCool
  Collaboration}}]{nicholson16}
{Nicholson}, B.~A., {Vidotto}, A.~A., {Mengel}, M., {et~al.} 2016, \mnras, 459,
  1907

\bibitem[{{{\'O} Fionnag{\'a}in} {et~al.}(2019){{\'O} Fionnag{\'a}in},
  {Vidotto}, {Petit}, {Folsom}, {Jeffers}, {Marsden}, {Morin}, {do Nascimento},
  \& {BCool Collaboration}}]{dualta19}
{{\'O} Fionnag{\'a}in}, D., {Vidotto}, A.~A., {Petit}, P., {et~al.} 2019,
  \mnras, 483, 873

\bibitem[{{Oran} {et~al.}(2017){Oran}, {Landi}, {van der Holst}, {Sokolov}, \&
  {Gombosi}}]{oran17}
{Oran}, R., {Landi}, E., {van der Holst}, B., {Sokolov}, I.~V., \& {Gombosi},
  T.~I. 2017, \apj, 845, 98

\bibitem[{{Oran} {et~al.}(2013){Oran}, {van der Holst}, {Landi}, {Jin},
  {Sokolov}, \& {Gombosi}}]{oran13}
{Oran}, R., {van der Holst}, B., {Landi}, E., {et~al.} 2013, \apj, 778, 176

\bibitem[{{Pagano} {et~al.}(2004){Pagano}, {Linsky}, {Valenti}, \&
  {Duncan}}]{pagano}
{Pagano}, I., {Linsky}, J.~L., {Valenti}, J., \& {Duncan}, D.~K. 2004, \aap,
  415, 331

\bibitem[{{Parker}(1958)}]{parker58}
{Parker}, E.~N. 1958, \apj, 128, 664

\bibitem[{{Parker}(1965)}]{parker65}
{Parker}, E.~N. 1965, \ssr, 4, 666

\bibitem[{{Pesnell} {et~al.}(2012){Pesnell}, {Thompson}, \&
  {Chamberlin}}]{pesnell12}
{Pesnell}, W.~D., {Thompson}, B.~J., \& {Chamberlin}, P.~C. 2012, \solphys,
  275, 3

\bibitem[{{Peter}(2006)}]{peter06}
{Peter}, H. 2006, \aap, 449, 759

\bibitem[{{Petit} {et~al.}(2008){Petit}, {Dintrans}, {Solanki}, {Donati},
  {Auri{\`e}re}, {Ligni{\`e}res}, {Morin}, {Paletou}, {Ramirez Velez},
  {Catala}, \& {Fares}}]{petit08}
{Petit}, P., {Dintrans}, B., {Solanki}, S.~K., {et~al.} 2008, \mnras, 388, 80

\bibitem[{{Petit} {et~al.}(2014){Petit}, {Louge}, {Th{\'e}ado}, {Paletou},
  {Manset}, {Morin}, {Marsden}, \& {Jeffers}}]{petit14}
{Petit}, P., {Louge}, T., {Th{\'e}ado}, S., {et~al.} 2014, \pasp, 126, 469

\bibitem[{{Pevtsov} {et~al.}(2003){Pevtsov}, {Fisher}, {Acton}, {Longcope},
  {Johns-Krull}, {Kankelborg}, \& {Metcalf}}]{pevtsov03}
{Pevtsov}, A.~A., {Fisher}, G.~H., {Acton}, L.~W., {et~al.} 2003, \apj, 598,
  1387

\bibitem[{{Phillips} {et~al.}(2008){Phillips}, {Feldman}, \&
  {Landi}}]{phillips08}
{Phillips}, K.~J.~H., {Feldman}, U., \& {Landi}, E. 2008, {Ultraviolet and
  X-ray Spectroscopy of the Solar Atmosphere} (Cambridge University Press)

\bibitem[{{Piskunov} \& {Kochukhov}(2002)}]{piskunov02}
{Piskunov}, N. \& {Kochukhov}, O. 2002, \aap, 381, 736

\bibitem[{{Pizzo} {et~al.}(1983){Pizzo}, {Schwenn}, {Marsch}, {Rosenbauer},
  {Muehlhaeuser}, \& {Neubauer}}]{pizzo83}
{Pizzo}, V., {Schwenn}, R., {Marsch}, E., {et~al.} 1983, \apj, 271, 335

\bibitem[{{Porto de Mello} \& {da Silva}(1997)}]{porto97}
{Porto de Mello}, G.~F. \& {da Silva}, L. 1997, \apjl, 482, L89

\bibitem[{{Powell} {et~al.}(1999){Powell}, {Roe}, {Linde}, {Gombosi}, \& {De
  Zeeuw}}]{powell99}
{Powell}, K.~G., {Roe}, P.~L., {Linde}, T.~J., {Gombosi}, T.~I., \& {De Zeeuw},
  D.~L. 1999, Journal of Computational Physics, 154, 284

\bibitem[{{R{\'e}ville} \& {Brun}(2017)}]{reville17}
{R{\'e}ville}, V. \& {Brun}, A.~S. 2017, \apj, 850, 45

\bibitem[{{R{\'e}ville} {et~al.}(2015){R{\'e}ville}, {Brun}, {Strugarek},
  {Matt}, {Bouvier}, {Folsom}, \& {Petit}}]{reville15}
{R{\'e}ville}, V., {Brun}, A.~S., {Strugarek}, A., {et~al.} 2015, \apj, 814, 99

\bibitem[{{Riley} {et~al.}(2014){Riley}, {Ben-Nun}, {Linker}, {Mikic},
  {Svalgaard}, {Harvey}, {Bertello}, {Hoeksema}, {Liu}, \& {Ulrich}}]{riley14}
{Riley}, P., {Ben-Nun}, M., {Linker}, J.~A., {et~al.} 2014, \solphys, 289, 769

\bibitem[{{Robinson} {et~al.}(1998){Robinson}, {Carpenter}, \&
  {Brown}}]{robinson98}
{Robinson}, R.~D., {Carpenter}, K.~G., \& {Brown}, A.~e. 1998, \apj, 503, 396

\bibitem[{{Ros{\'e}n} {et~al.}(2016){Ros{\'e}n}, {Kochukhov}, {Hackman}, \&
  {Lehtinen}}]{rosen16}
{Ros{\'e}n}, L., {Kochukhov}, O., {Hackman}, T., \& {Lehtinen}, J. 2016, \aap,
  593, A35

\bibitem[{{Sachdeva} {et~al.}(2019){Sachdeva}, {van der Holst}, {Manchester},
  {T{\'o}th}, {Chen}, {Lloveras}, {V{\'a}squez}, {Lamy}, {Wojak}, {Jackson},
  {Yu}, \& {Henney}}]{sachdeva19}
{Sachdeva}, N., {van der Holst}, B., {Manchester}, W.~B., {et~al.} 2019, arXiv
  e-prints, arXiv:1910.08110

\bibitem[{{See} {et~al.}(2018){See}, {Jardine}, {Vidotto}, {Donati}, {Boro
  Saikia}, {Fares}, {Folsom}, {Jeffers}, {Marsden}, {Morin}, {Petit}, \& {BCool
  Collaboration}}]{see18}
{See}, V., {Jardine}, M., {Vidotto}, A.~A., {et~al.} 2018, \mnras, 474, 536

\bibitem[{{See} {et~al.}(2019){See}, {Matt}, {Finley}, {Folsom}, {Boro Saikia},
  {Donati}, {Fares}, {H{\'e}brard}, {Jardine}, {Jeffers}, {Marsden}, {Mengel},
  {Morin}, {Petit}, {Vidotto}, {Waite}, \& {the BCool Collaboration}}]{see19}
{See}, V., {Matt}, S.~P., {Finley}, A.~J., {et~al.} 2019, \apj, 886, 120

\bibitem[{{Semel}(1989)}]{semel89}
{Semel}, M. 1989, \aap, 225, 456

\bibitem[{{Sokolov} {et~al.}(2013){Sokolov}, {van der Holst}, {Oran}, {Downs},
  {Roussev}, {Jin}, {Manchester}, {Evans}, \& {Gombosi}}]{sokolov13}
{Sokolov}, I.~V., {van der Holst}, B., {Oran}, R., {et~al.} 2013, \apj, 764, 23

\bibitem[{{Spitzer}(1956)}]{spitzer56}
{Spitzer}, L. 1956, {Physics of Fully Ionized Gases}

\bibitem[{{Stone} {et~al.}(1998){Stone}, {Frandsen}, {Mewaldt}, {Christian},
  {Margolies}, {Ormes}, \& {Snow}}]{stone98}
{Stone}, E.~C., {Frandsen}, A.~M., {Mewaldt}, R.~A., {et~al.} 1998, \ssr, 86, 1

\bibitem[{{Suzuki} {et~al.}(2013){Suzuki}, {Imada}, {Kataoka}, {Kato},
  {Matsumoto}, {Miyahara}, \& {Tsuneta}}]{suzuki13}
{Suzuki}, T.~K., {Imada}, S., {Kataoka}, R., {et~al.} 2013, \pasj, 65, 98

\bibitem[{{Suzuki} \& {Inutsuka}(2006)}]{suzuki06}
{Suzuki}, T.~K. \& {Inutsuka}, S.-I. 2006, Journal of Geophysical Research
  (Space Physics), 111, A06101

\bibitem[{{Tarduno} {et~al.}(2010){Tarduno}, {Cottrell}, {Watkeys}, {Hofmann},
  {Doubrovine}, {Mamajek}, {Liu}, {Sibeck}, {Neukirch}, \& {Usui}}]{Tarduno10}
{Tarduno}, J.~A., {Cottrell}, R.~D., {Watkeys}, M.~K., {et~al.} 2010, Science,
  327, 1238

\bibitem[{{Tian} {et~al.}(2008){Tian}, {Kasting}, {Liu}, \& {Roble}}]{tian08}
{Tian}, F., {Kasting}, J.~F., {Liu}, H.-L., \& {Roble}, R.~G. 2008, Journal of
  Geophysical Research (Planets), 113, E05008

\bibitem[{{T{\'o}th} {et~al.}(2011){T{\'o}th}, {van der Holst}, \&
  {Huang}}]{toth2011}
{T{\'o}th}, G., {van der Holst}, B., \& {Huang}, Z. 2011, \apj, 732, 102

\bibitem[{{T{\'o}th} {et~al.}(2012){T{\'o}th}, {van der Holst}, {Sokolov}, {De
  Zeeuw}, {Gombosi}, {Fang}, {Manchester}, {Meng}, {Najib}, {Powell}, {Stout},
  {Glocer}, {Ma}, \& {Opher}}]{toth12}
{T{\'o}th}, G., {van der Holst}, B., {Sokolov}, I.~V., {et~al.} 2012, Journal
  of Computational Physics, 231, 870

\bibitem[{{Usmanov} {et~al.}(2000){Usmanov}, {Goldstein}, {Besser}, \&
  {Fritzer}}]{usmanov00}
{Usmanov}, A.~V., {Goldstein}, M.~L., {Besser}, B.~P., \& {Fritzer}, J.~M.
  2000, \jgr, 105, 12675

\bibitem[{{Usmanov} {et~al.}(2018){Usmanov}, {Matthaeus}, {Goldstein}, \&
  {Chhiber}}]{usmanov18}
{Usmanov}, A.~V., {Matthaeus}, W.~H., {Goldstein}, M.~L., \& {Chhiber}, R.
  2018, \apj, 865, 25

\bibitem[{{Valenti} \& {Fischer}(2005)}]{valenti05}
{Valenti}, J.~A. \& {Fischer}, D.~A. 2005, \apjs, 159, 141

\bibitem[{{van den Oord} \& {Doyle}(1997)}]{vanoord97}
{van den Oord}, G.~H.~J. \& {Doyle}, J.~G. 1997, \aap, 319, 578

\bibitem[{{van der Holst} {et~al.}(2007){van der Holst}, {Jacobs}, \&
  {Poedts}}]{vanderholst07}
{van der Holst}, B., {Jacobs}, C., \& {Poedts}, S. 2007, \apjl, 671, L77

\bibitem[{{van der Holst} {et~al.}(2014){van der Holst}, {Sokolov}, {Meng},
  {Jin}, {Manchester}, {T{\'o}th}, \& {Gombosi}}]{vanderholst14}
{van der Holst}, B., {Sokolov}, I.~V., {Meng}, X., {et~al.} 2014, \apj, 782, 81

\bibitem[{{Vidotto}(2016)}]{vidotto16}
{Vidotto}, A.~A. 2016, \mnras, 459, 1533

\bibitem[{{Vidotto} \& {Bourrier}(2017)}]{vidotto17}
{Vidotto}, A.~A. \& {Bourrier}, V. 2017, \mnras, 470, 4026

\bibitem[{{Vidotto} {et~al.}(2014){Vidotto}, {Jardine}, {Morin}, {Donati},
  {Opher}, \& {Gombosi}}]{vidotto14}
{Vidotto}, A.~A., {Jardine}, M., {Morin}, J., {et~al.} 2014, \mnras, 438, 1162

\bibitem[{{Vidotto} {et~al.}(2011){Vidotto}, {Jardine}, {Opher}, {Donati}, \&
  {Gombosi}}]{vidotto11}
{Vidotto}, A.~A., {Jardine}, M., {Opher}, M., {Donati}, J.~F., \& {Gombosi},
  T.~I. 2011, \mnras, 412, 351

\bibitem[{{Villadsen} {et~al.}(2014){Villadsen}, {Hallinan}, {Bourke},
  {G{\"u}del}, \& {Rupen}}]{villadsen14}
{Villadsen}, J., {Hallinan}, G., {Bourke}, S., {G{\"u}del}, M., \& {Rupen}, M.
  2014, \apj, 788, 112

\bibitem[{{Wang} \& {Sheeley}(1990)}]{wangsheeley90}
{Wang}, Y.-M. \& {Sheeley}, Jr., N.~R. 1990, \apj, 355, 726

\bibitem[{{Wargelin} \& {Drake}(2002)}]{wargelin02}
{Wargelin}, B.~J. \& {Drake}, J.~J. 2002, \apj, 578, 503

\bibitem[{{Weber} \& {Davis}(1967)}]{weber67}
{Weber}, E.~J. \& {Davis}, Jr., L. 1967, \apj, 148, 217

\bibitem[{{Wood}(2004)}]{wood04}
{Wood}, B.~E. 2004, Living Reviews in Solar Physics, 1, 2

\bibitem[{{Wood} {et~al.}(1997){Wood}, {Linsky}, \& {Ayres}}]{wood97}
{Wood}, B.~E., {Linsky}, J.~L., \& {Ayres}, T.~R. 1997, \apj, 478, 745

\bibitem[{{Wood} {et~al.}(2001){Wood}, {Linsky}, {M{\"u}ller}, \&
  {Zank}}]{wood01}
{Wood}, B.~E., {Linsky}, J.~L., {M{\"u}ller}, H.-R., \& {Zank}, G.~P. 2001,
  \apjl, 547, L49

\bibitem[{{Wood} {et~al.}(2002){Wood}, {M{\"u}ller}, {Zank}, \&
  {Linsky}}]{wood02}
{Wood}, B.~E., {M{\"u}ller}, H.-R., {Zank}, G.~P., \& {Linsky}, J.~L. 2002,
  \apj, 574, 412

\bibitem[{{Wood} {et~al.}(2005){Wood}, {M{\"u}ller}, {Zank}, {Linsky}, \&
  {Redfield}}]{wood05}
{Wood}, B.~E., {M{\"u}ller}, H.-R., {Zank}, G.~P., {Linsky}, J.~L., \&
  {Redfield}, S. 2005, \apjl, 628, L143

\end{thebibliography}
\begin{appendix}
\begin{figure}
\section{Gaussian fit}
%Non-thermal velocity}
\centering
\includegraphics[width=.5\textwidth]{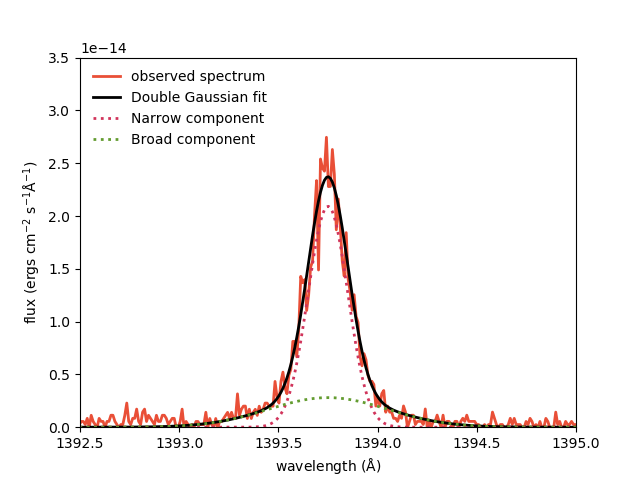}
\caption{Double Gaussian fit of the \ion{Si}{IV} line at 1393.75 \AA.}
\label{si}
\end{figure}
\begin{figure}
\section{Mass and angular momentum loss rates versus distance}
\centering
\includegraphics[width=.5\textwidth]{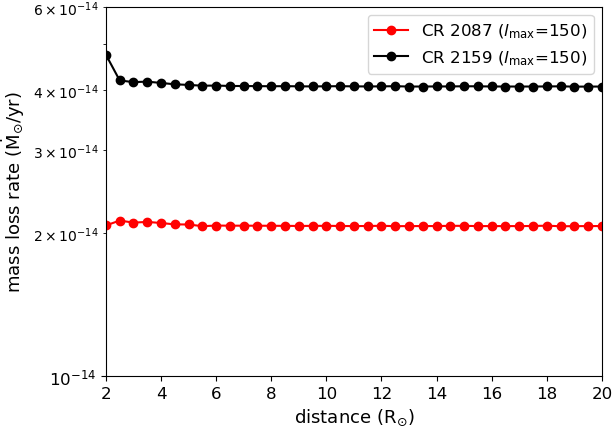} \\
\includegraphics[width=.5\textwidth]{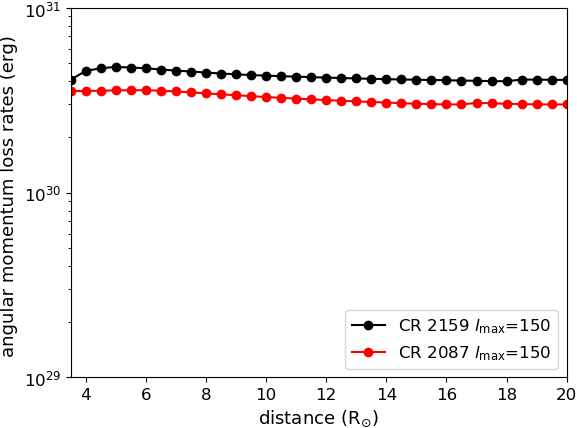}
\caption{Mass loss (\textit{top}) and angular momentum loss (\textit{bottom}) rates are shown as a function of radius 
for both solar maximum (CR 2159, black) and solar minimum case (CR 2087, red).}
\label{mdotjdot}
\end{figure}
\begin{figure*}
\section{Number density versus velocity $u_\mathrm{r}$}
\centering
\includegraphics[width=1.\textwidth]{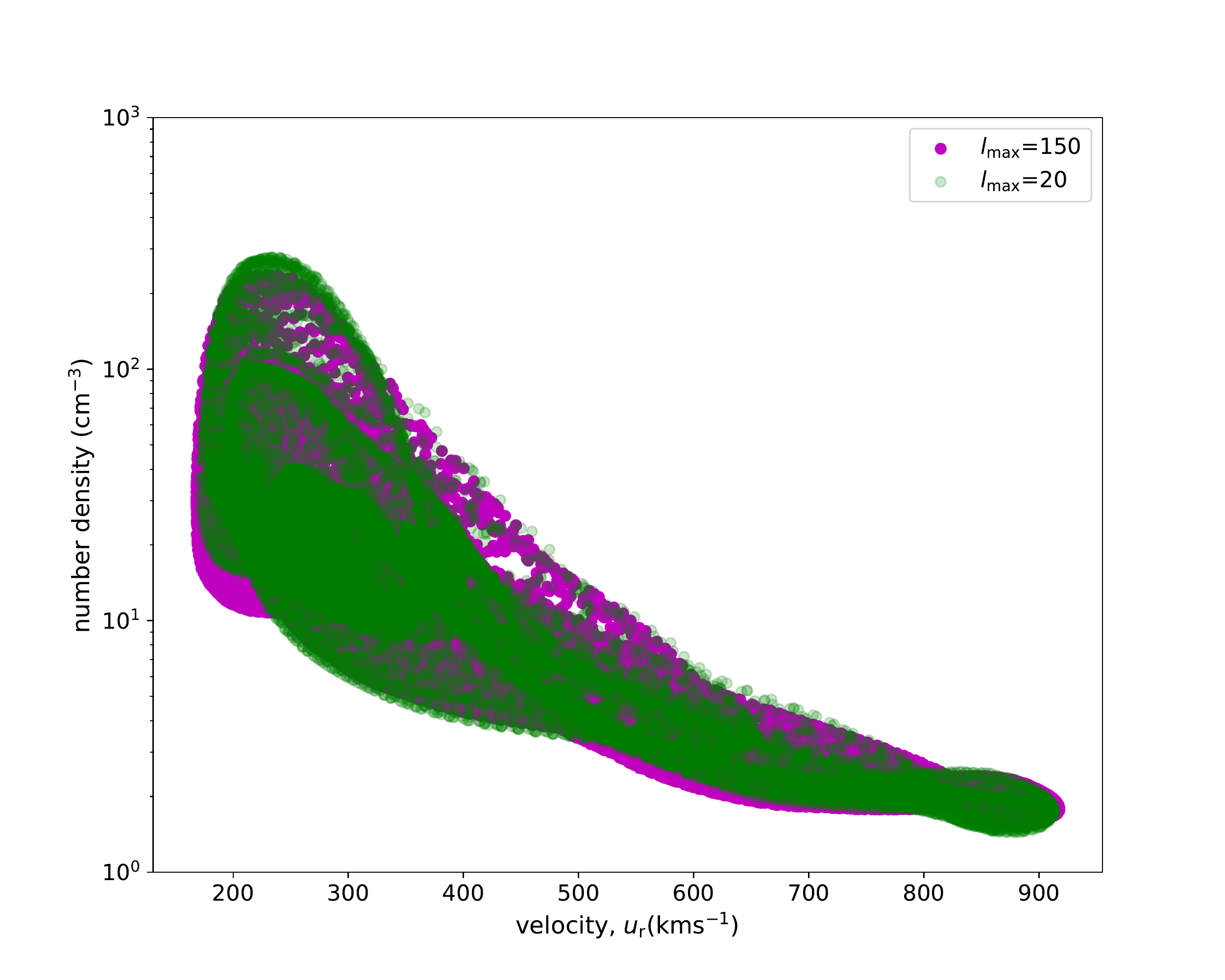}
\caption{Proton number density vs. wind velocity, $u_\mathrm{r}$ during solar maximum, CR 2159. The $l_\mathrm{max}$=150 
simulation is shown in magenta and $l_\mathrm{max}$=20 simulation output is shown in green.}
\label{rhour}
\end{figure*}

\begin{figure*}
\section{Solar wind speed, proton density, and ram pressure for $l_\mathrm{max}$=150 and 5 over the given range of $S_\mathrm{A}/B$}
\centering
\includegraphics[width=.34\textwidth]{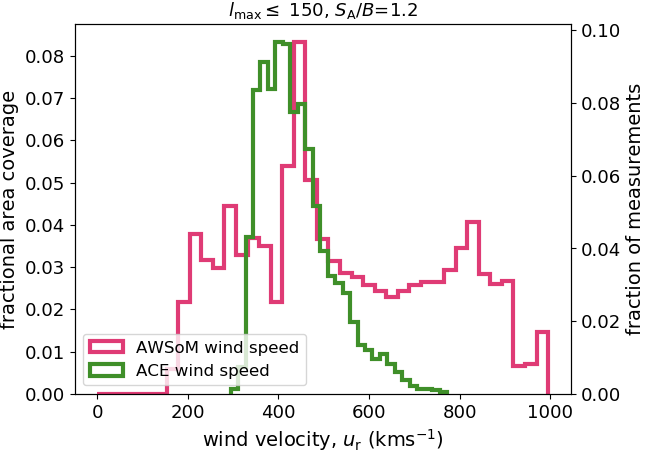}~~~\includegraphics[width=.34\textwidth]{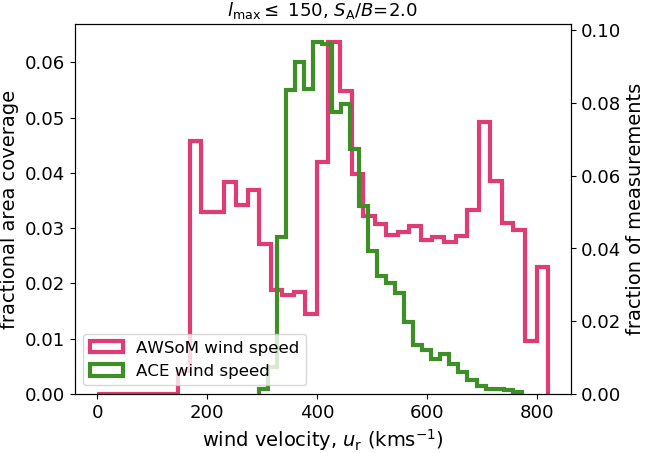}~~~\includegraphics[width=.34\textwidth]{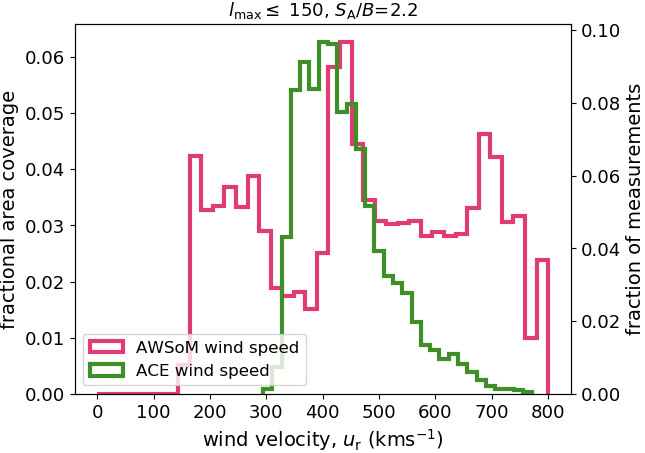}\\
\includegraphics[width=.34\textwidth]{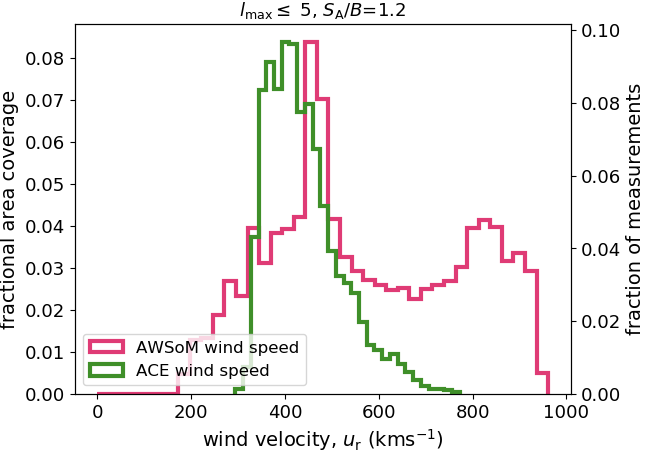}~~~\includegraphics[width=.34\textwidth]{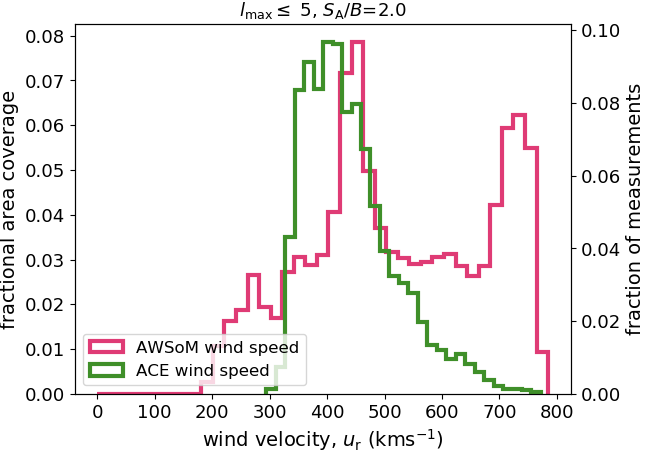}~~~\includegraphics[width=.34\textwidth]{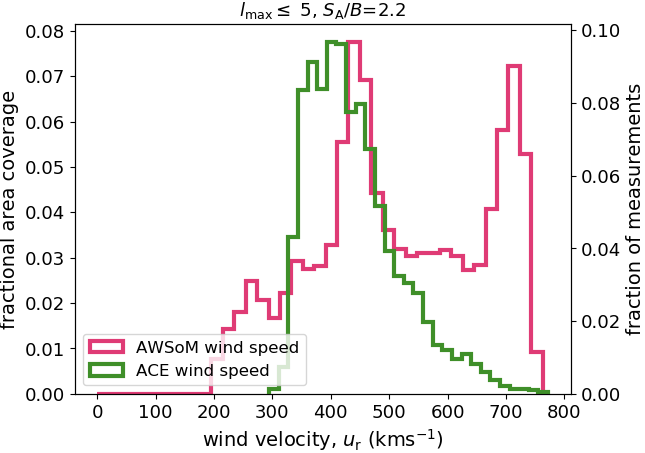}\\
\caption{Distribution of wind velocity at 1 AU for a subset of our simulations during solar cycle minimum
and maximum. Each column represents a steady state simulation for different values of $S_\mathrm{A}/B$. 
The resolution of the magnetogram is truncated to $l_\mathrm{max}$=150 (\textit{top}) and $l_\mathrm{max}$=5
(\textit{bottom}). ACE data is shown in green.}
\label{vgridmax}
\end{figure*}

\begin{figure*}
\centering
\includegraphics[width=.34\textwidth]{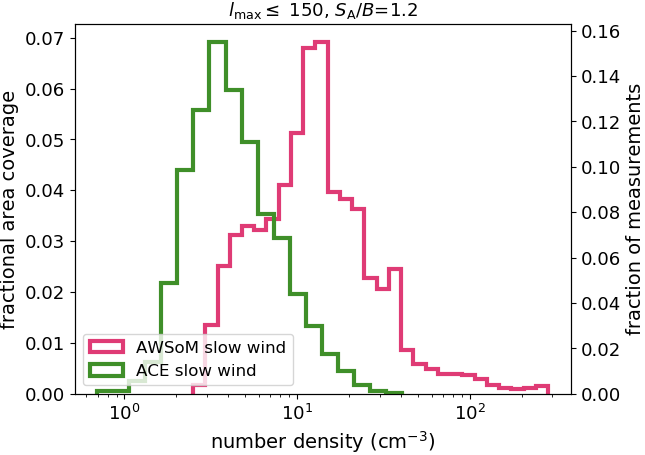}~~~\includegraphics[width=.34\textwidth]{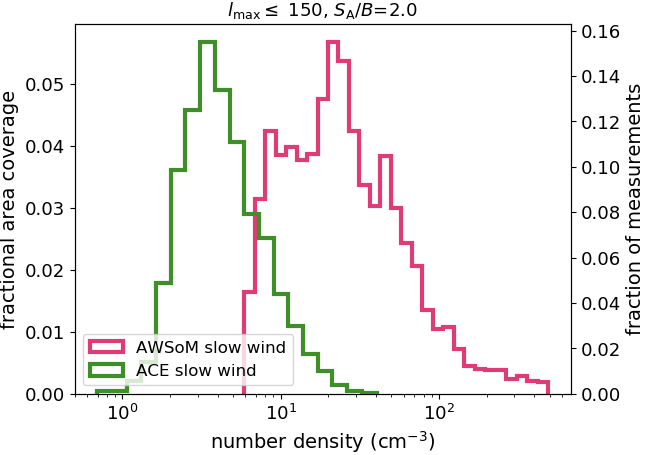}~~~\includegraphics[width=.34\textwidth]{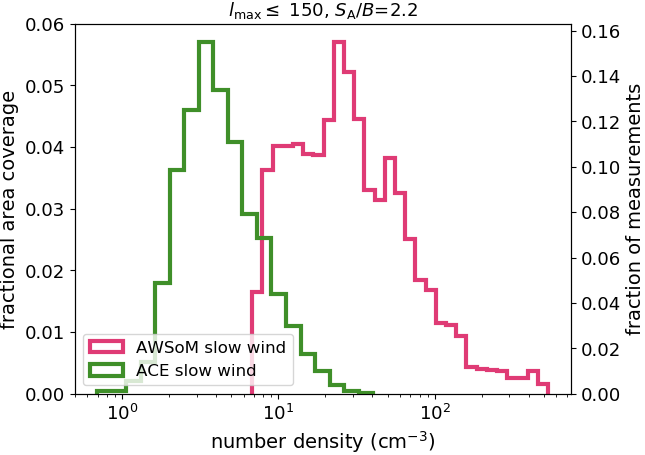}\\
\includegraphics[width=.34\textwidth]{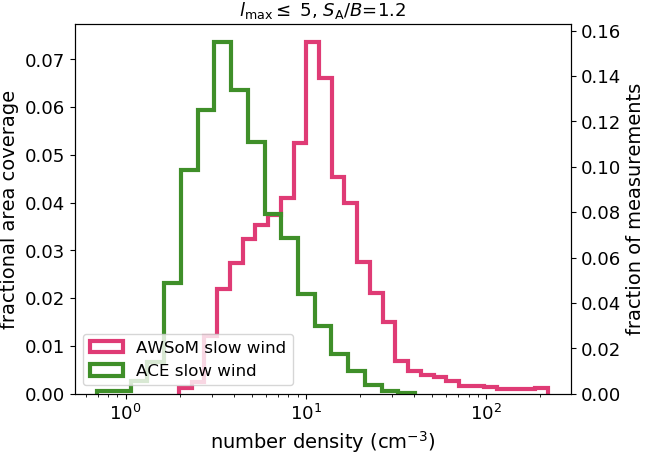}~~~\includegraphics[width=.34\textwidth]{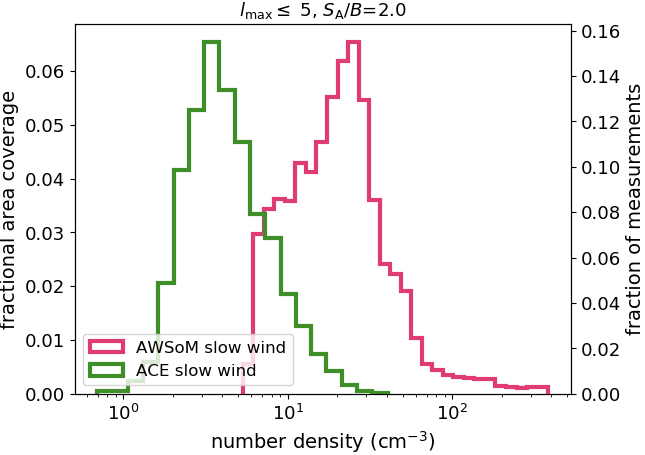}~~~\includegraphics[width=.34\textwidth]{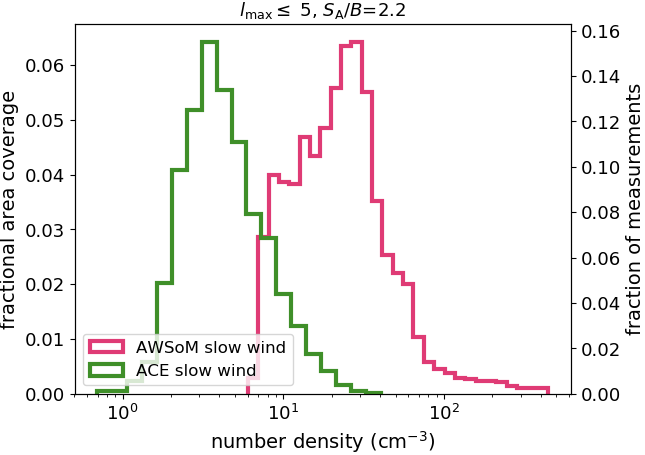}\\
\caption{Proton number density of the slow wind for the same simulations as in Fig .\ref{vgridmax}. The observed ACE proton
densities are shown in green.}
\label{rhogridslow}
\end{figure*}

\begin{figure*}
\centering
\includegraphics[width=.34\textwidth]{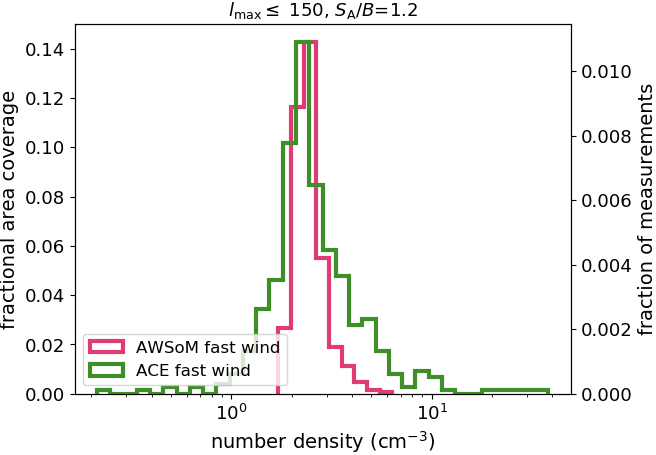}~~~\includegraphics[width=.34\textwidth]{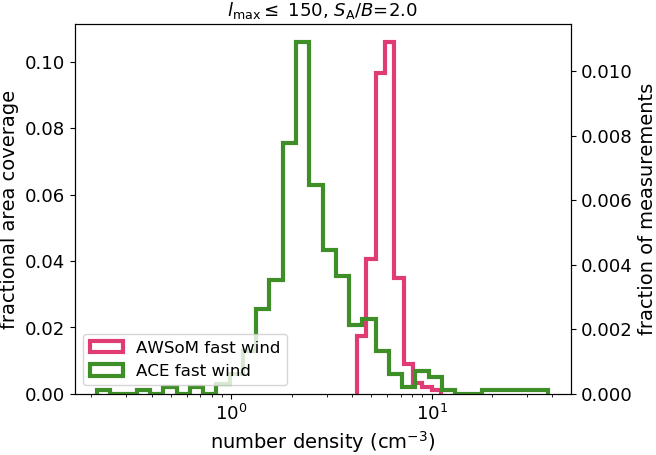}~~~\includegraphics[width=.34\textwidth]{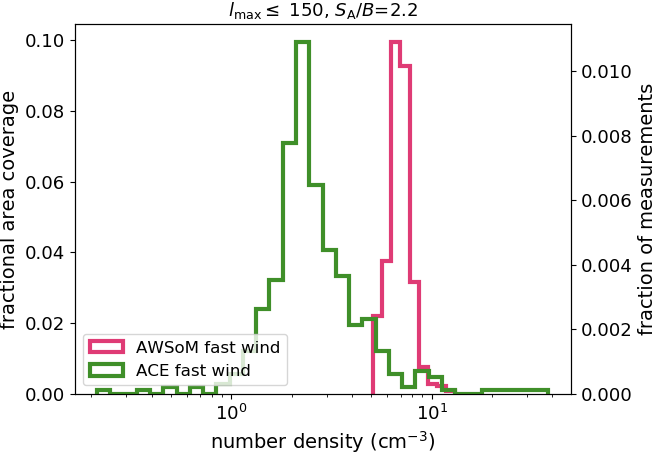}\\
\includegraphics[width=.34\textwidth]{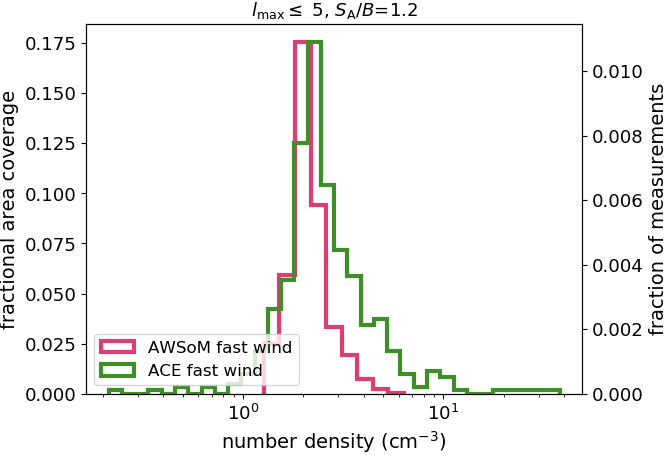}~~~\includegraphics[width=.34\textwidth]{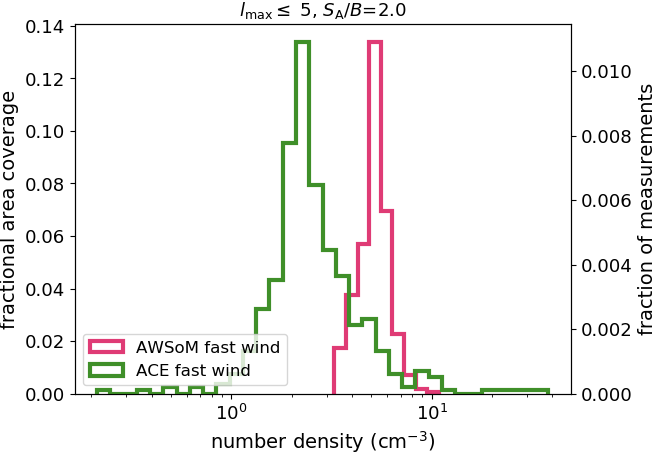}~~~\includegraphics[width=.34\textwidth]{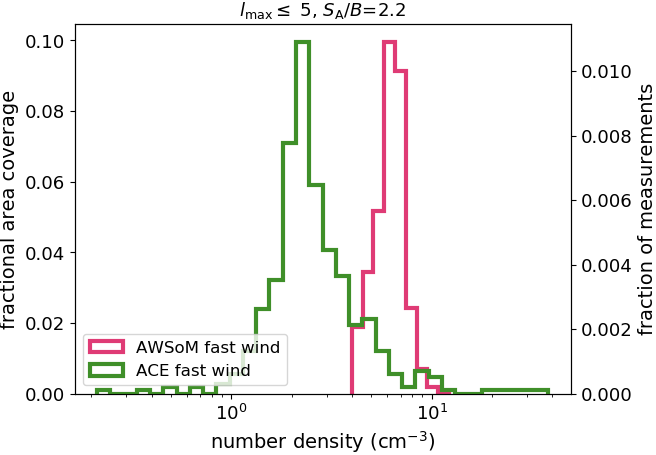}\\
\caption{Same as in Fig. \ref{rhogridslow} but for the fast wind.}
\label{rhogridfast}
\end{figure*}

\begin{figure*}
\centering
\includegraphics[width=.34\textwidth]{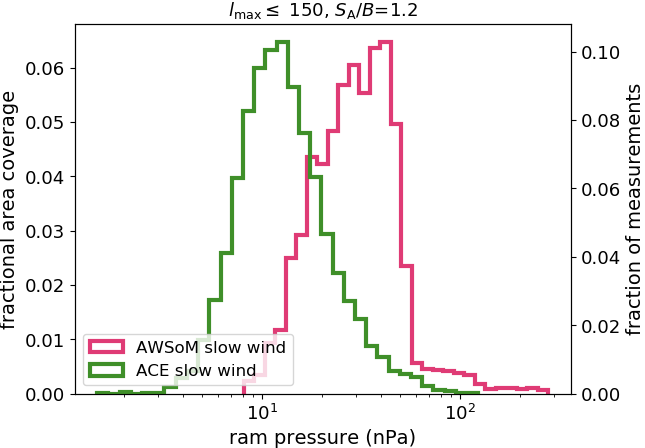}~~~\includegraphics[width=.34\textwidth]{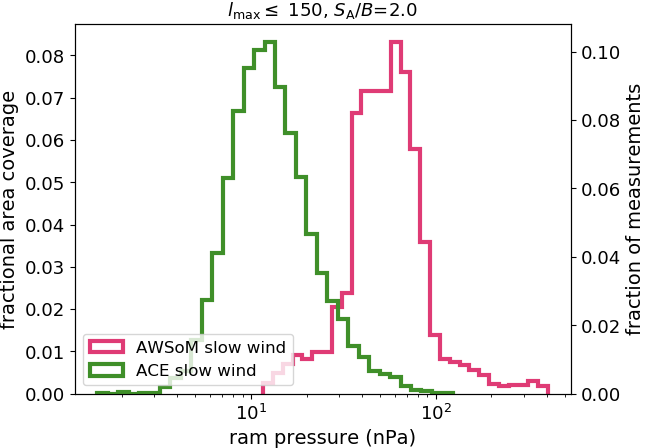}~~~\includegraphics[width=.34\textwidth]{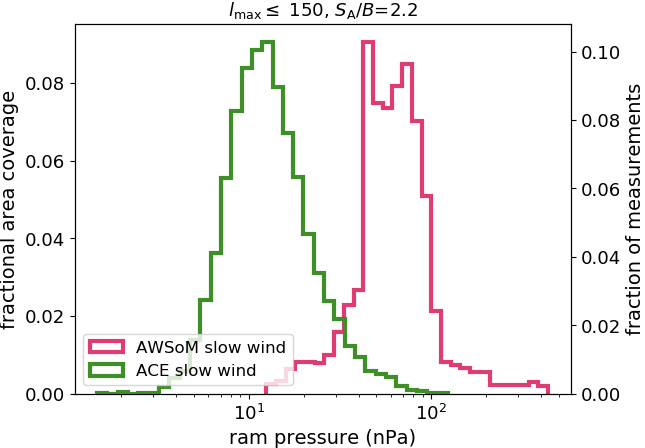}\\
\includegraphics[width=.34\textwidth]{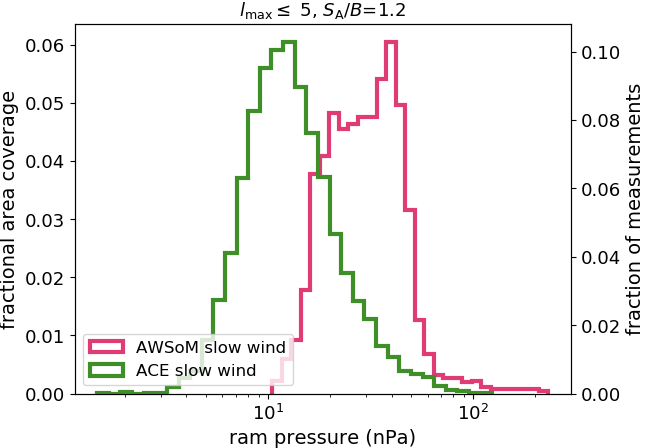}~~~\includegraphics[width=.34\textwidth]{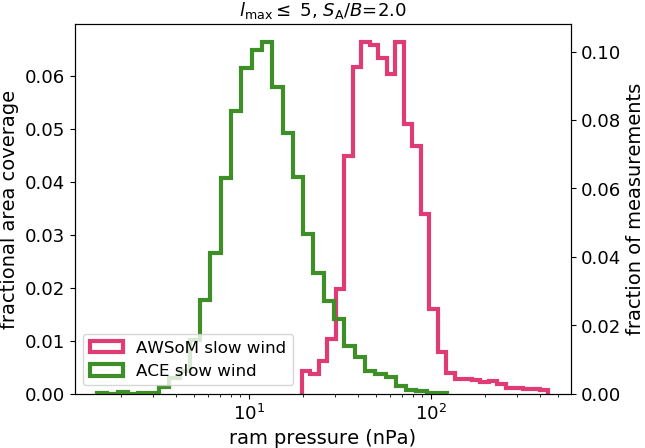}~~~\includegraphics[width=.34\textwidth]{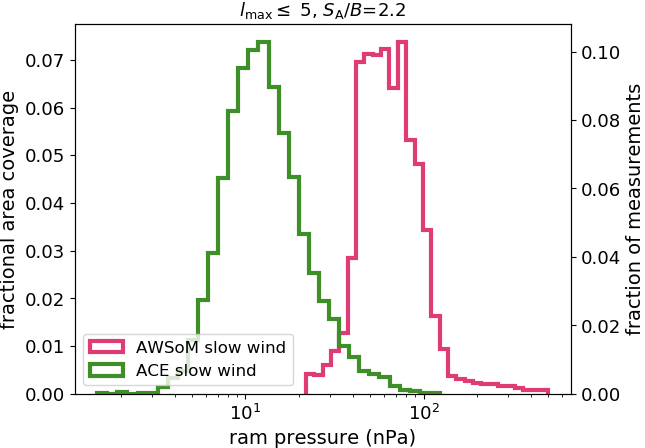}\\
\caption{Ram pressure for the same simulations as in Fig. \ref{vgridmax}. ACE ram pressure is shown in green.}
\label{ramgridslow}
\end{figure*}

\begin{figure*}
\centering
\includegraphics[width=.34\textwidth]{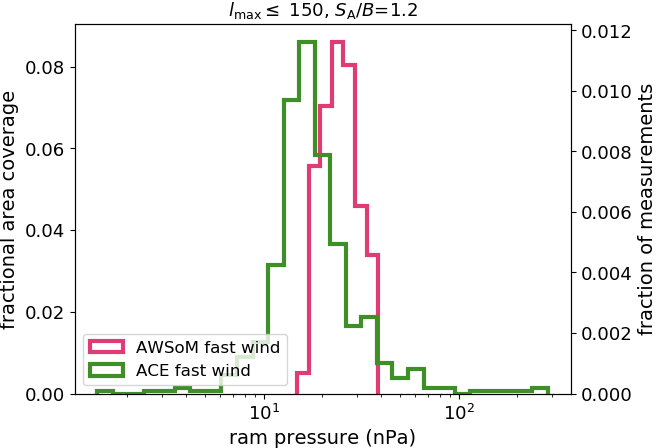}~~~\includegraphics[width=.34\textwidth]{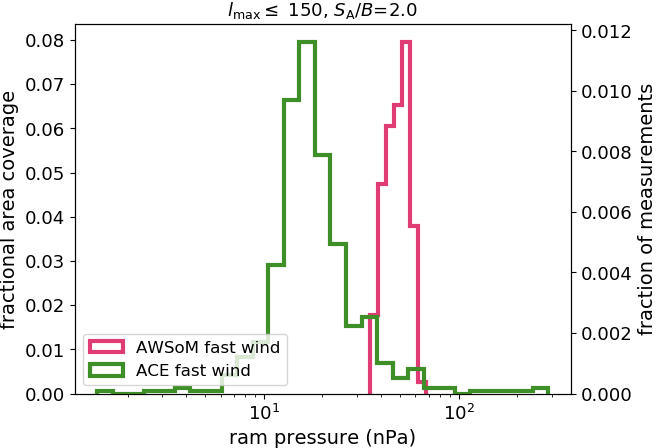}~~~\includegraphics[width=.34\textwidth]{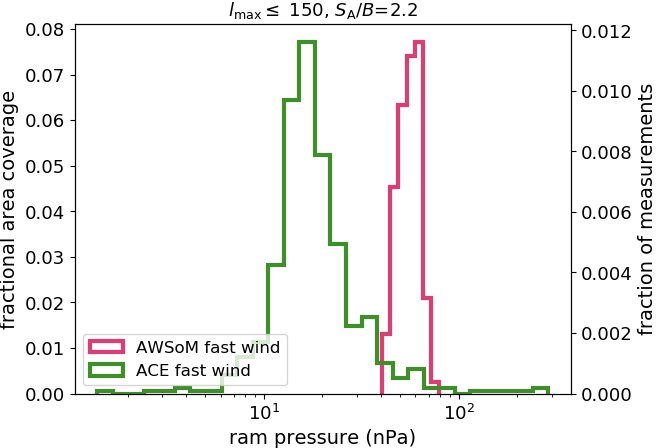}\\
\includegraphics[width=.34\textwidth]{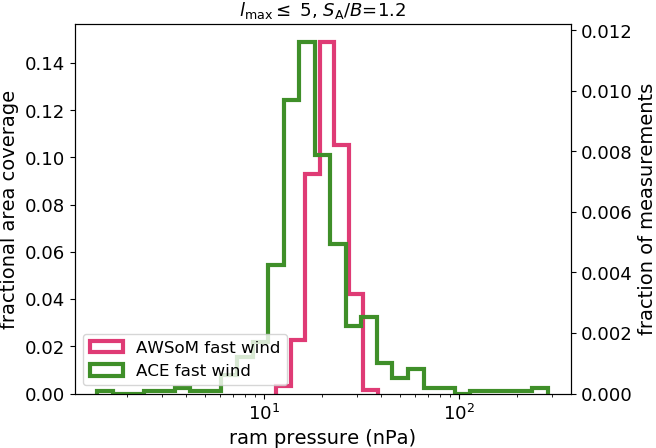}~~~\includegraphics[width=.34\textwidth]{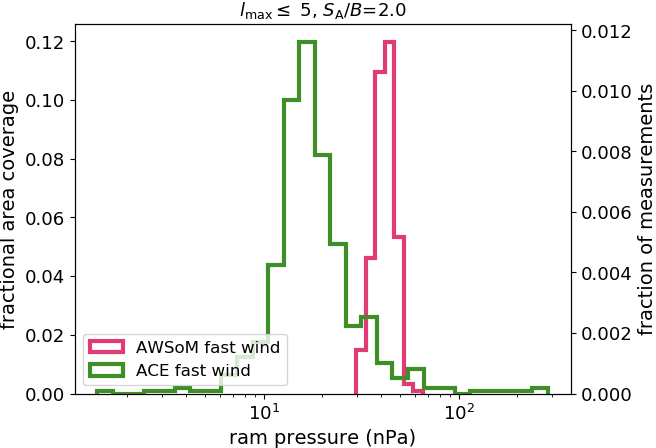}~~~\includegraphics[width=.34\textwidth]{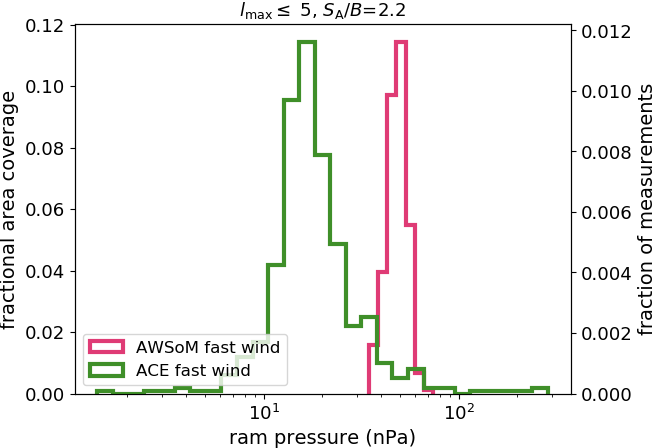}\\
\caption{Same as in Fig. \ref{ramgridslow} but for the fast wind.}
\label{ramgridfast}
\end{figure*}

\begin{figure}
\section{ZDI solar simulations}
\includegraphics[width=.5\textwidth]{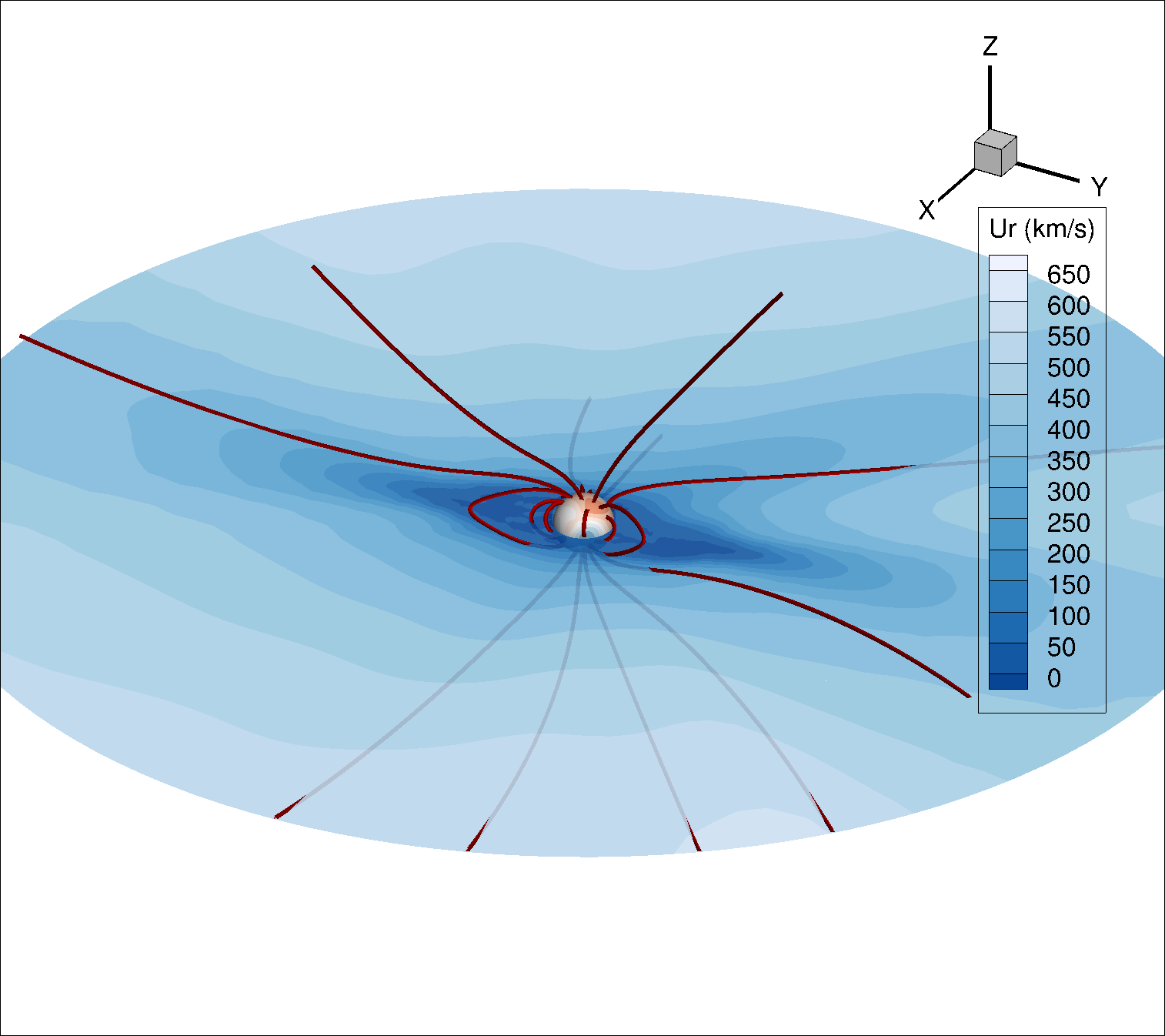}
\caption{Same as Fig.~\ref{3dplot} but for HN Peg.}
\label{3dplothnpeg}
\end{figure}
\begin{figure*}
\centering
\includegraphics[width=.33\textwidth]{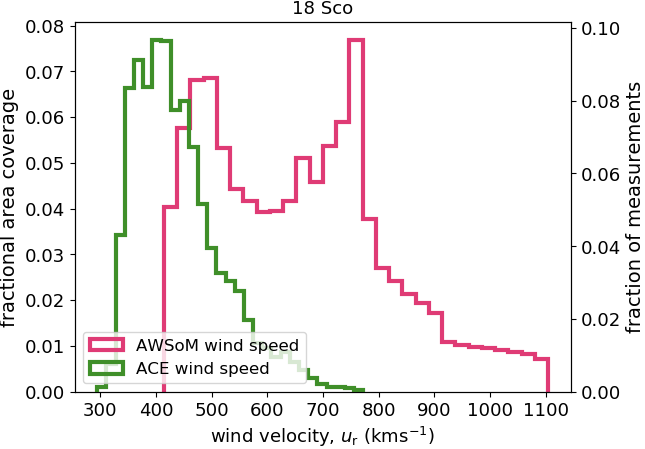}~~~\includegraphics[width=.33\textwidth]{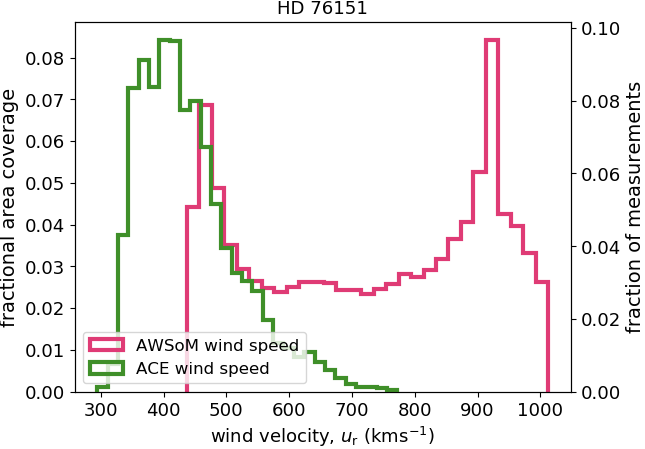}~~~\includegraphics[width=.33\textwidth]{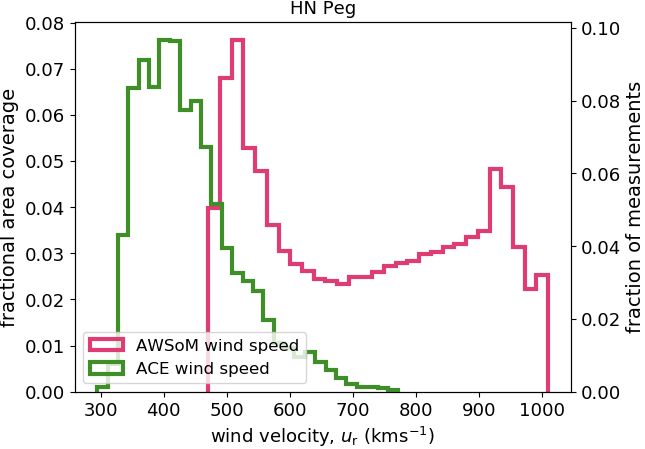}
\caption{Wind velocity determined from ZDI simulations of the three stars: 18 Sco, HD 76151, HN Peg (\textit{left} to \textit{right}) in magenta. The observed ACE wind velocities are shown in green.}
\label{urzdi}
\end{figure*}

\begin{figure*}
\centering
\includegraphics[width=.33\textwidth]{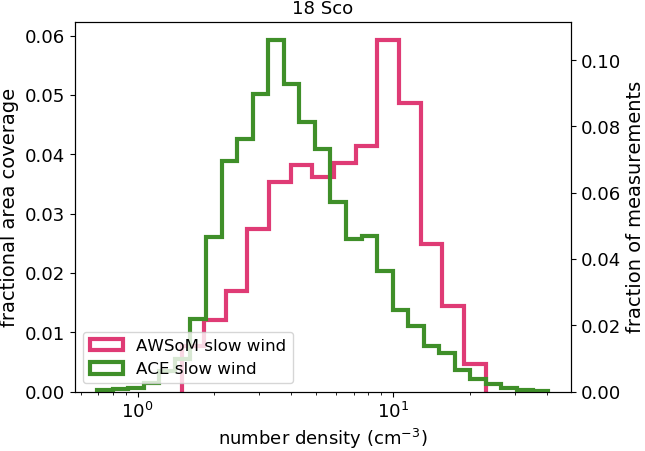}~~~\includegraphics[width=.33\textwidth]{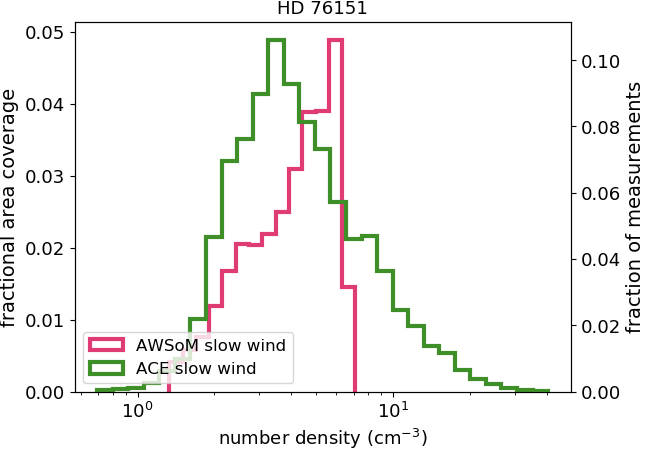}~~~\includegraphics[width=.33\textwidth]{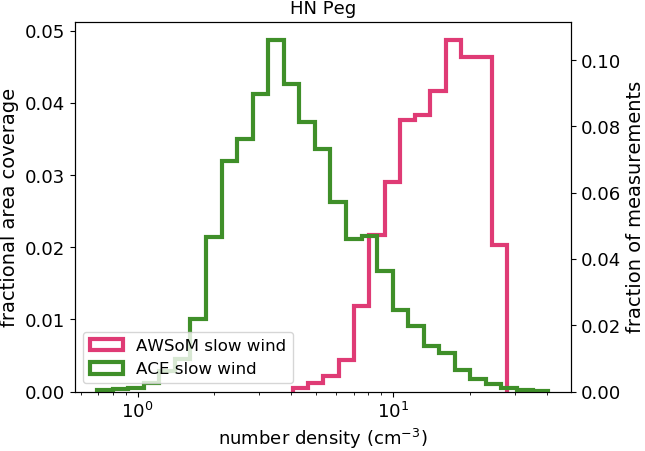}\\
\includegraphics[width=.33\textwidth]{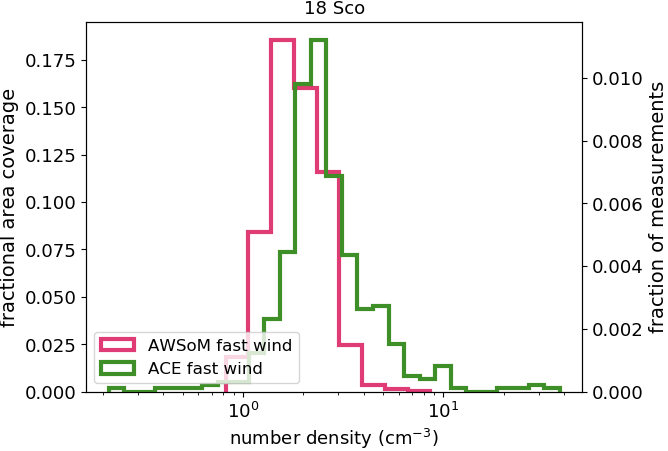}~~~\includegraphics[width=.33\textwidth]{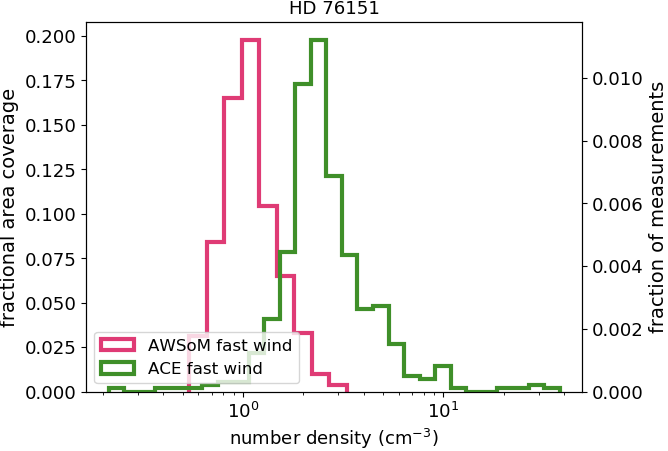}~~~\includegraphics[width=.33\textwidth]{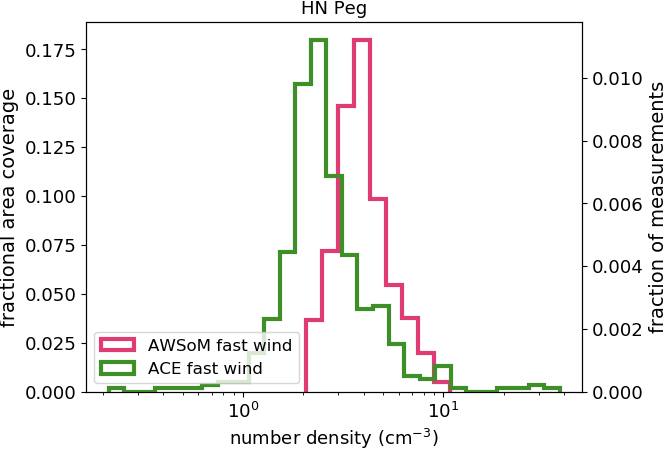}
\caption{Proton number density for the slow (\textit{top}) and fast (\textit{bottom}) 
component of the wind. Each column represents wind simulations for the three stars 
included in this work: 18 Sco, HD76151, HN Peg (\textit{left} to \textit{right}, magenta). ACE proton density is shown in green.}
\label{rhozdi}
\end{figure*}
\begin{figure*}
\centering
\includegraphics[width=.33\textwidth]{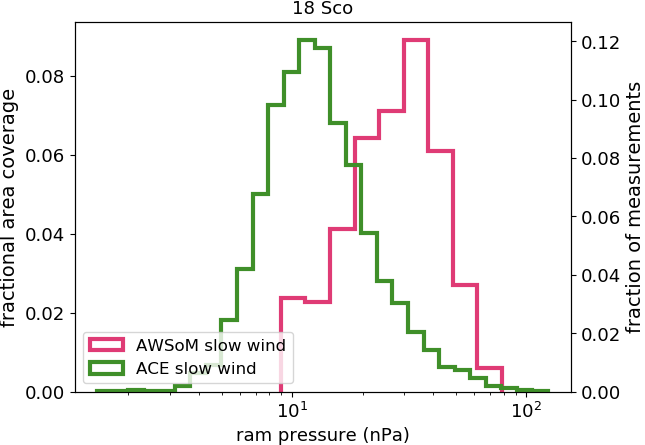}~~~\includegraphics[width=.33\textwidth]{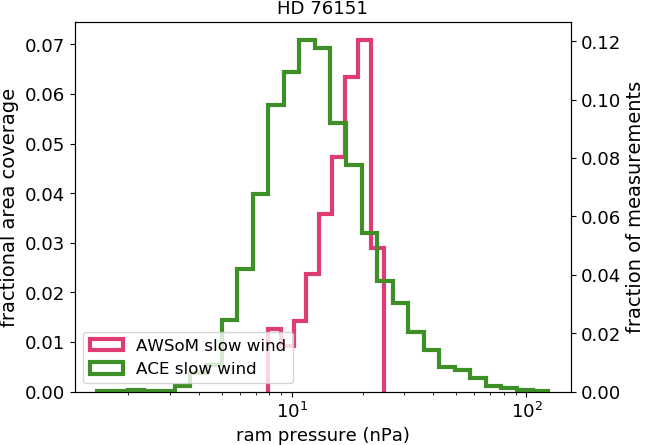}~~~\includegraphics[width=.33\textwidth]{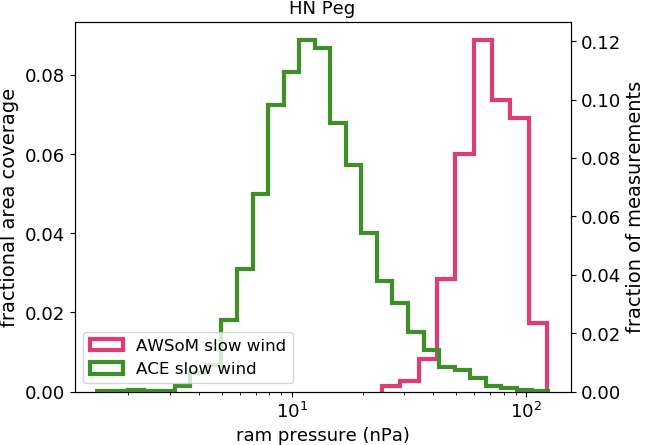}\\
\includegraphics[width=.33\textwidth]{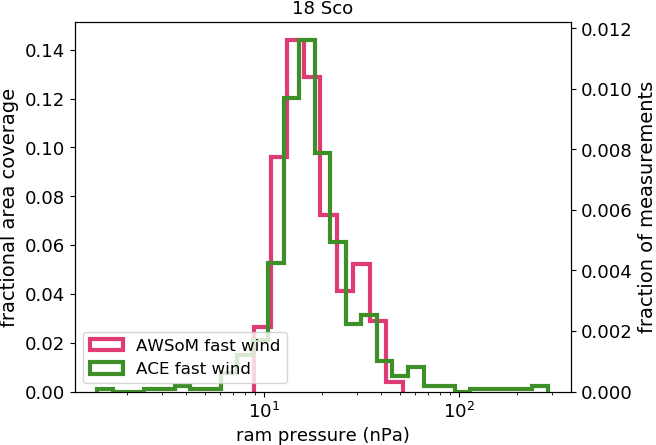}~~~\includegraphics[width=.33\textwidth]{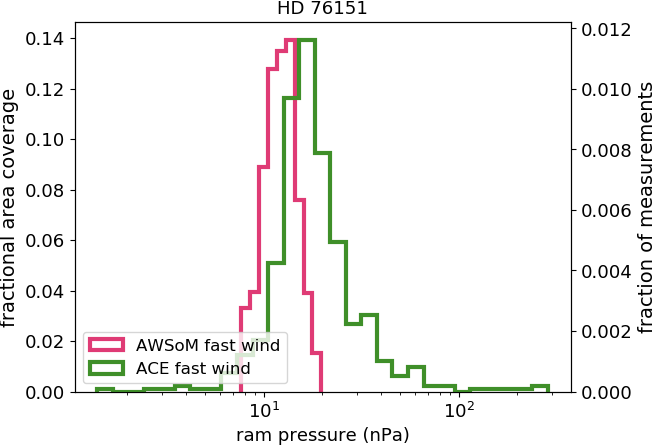}~~~\includegraphics[width=.33\textwidth]{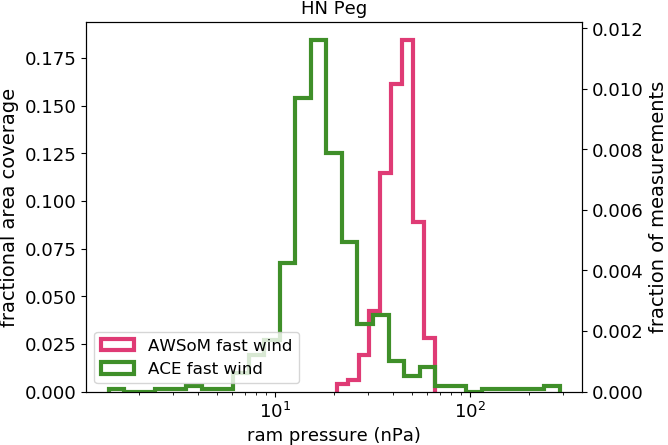}
\caption{Same as in Fig. \ref{rhozdi} except the ram pressure is shown instead of the proton density.}
\label{ramzdi}
\end{figure*}
%%%%%
\begin{table*}
\section{Tables}
\caption{Mass loss and angular momentum loss 
rates during solar cycle maximum (CR 2159) 
for different values of 
$l_\mathrm{max}$ and $S_\mathrm{A}/B \times 10^6$ W~m$^{-2}$~T$^{-1}$ in the grid. }
\label{gridtable}
\centering
\begin{tabular}{|c|c|c|c|c|c|c|c|c|}
\hline
\multirow{2}{*}&
\multicolumn{4}{c}{$\dot{M}, \mathrm{\times 10^{-14}} \mathrm{\dot{M}_\sun} \mathrm{yr^{-1}}$} &
\multicolumn{4}{|c|}{$\dot{J}, \mathrm{\times 10^{30}} $erg}\\\cline{2-9}
&$S_\mathrm{A}/B$=1.1&$S_\mathrm{A}/B$=1.2&$S_\mathrm{A}/B$=2.0&$S_\mathrm{A}/B$=2.2&$S_\mathrm{A}/B$=1.1&$S_\mathrm{A}/B$=1.2&$S_\mathrm{A}/B$=2.0&$S_\mathrm{A}/B$=2.2\\
\hline
{$l_\mathrm{max}=5$}&2.9&3.2&6.7&7.7&2.8&2.9&4.8&5.4 \\
\hline
{$l_\mathrm{max}=10$}&3.4&3.8&8.2&9.6&3.2&3.5&6.0&6.8\\
\hline
{$l_\mathrm{max}=20$}&4.4&4.2&8.9&10.3&4.2&3.7&7.0&8.0\\
\hline
{$l_\mathrm{max}=150$}&4.1&3.9&8.4&9.7&4.0&3.7&6.5&7.3\\
\hline
\end{tabular}
\end{table*}
%%%%%%%%%
%%%%%%%%
%%%%%%%%%%
%%%%%%%%%%
\begin{table*}
\caption{Same as Table \ref{gridtable} but during solar minimum (CR 2087).}
\label{gridtable2}
\centering
\begin{tabular}{|c|c|c|c|c|c|c|c|c|}
\hline
\multirow{2}{*}&
\multicolumn{4}{c}{$\dot{M}, \mathrm{\times 10^{-14}} \mathrm{\dot{M}_\sun} \mathrm{yr^{-1}}$} &
\multicolumn{4}{|c|}{$\dot{J}, \mathrm{\times 10^{30}} $erg}\\\cline{2-9}
&$S_\mathrm{A}/B$=1.1&$S_\mathrm{A}/B$=1.2&$S_\mathrm{A}/B$=2.0&$S_\mathrm{A}/B$=2.2&$S_\mathrm{A}/B$=1.1&$S_\mathrm{A}/B$=1.2&$S_\mathrm{A}/B$=2.0&$S_\mathrm{A}/B$=2.2\\
\hline
{$l_\mathrm{max}=5$}&1.7&2.1&4.0&4.6&2.4&2.5&4.1&4.5 \\
\hline
{$l_\mathrm{max}=10$}&1.8&2.0&4.1&4.7&3.0&2.7&4.2&4.6\\
\hline
{$l_\mathrm{max}=20$}&2.0&2.0&4.1&4.7&3.0&2.6&4.2&4.7\\
\hline
{$l_\mathrm{max}=150$}&2.1&2.1&4.2&4.9&3.0&2.8&4.3&4.7\\
\hline
\end{tabular}
\end{table*}

%%%%%
\begin{table*}
\caption{Median and mean values of the wind speed, proton density, and ram pressure for the slow and the fast wind 
for simulations where $l_\mathrm{max}$=150, and $S_\mathrm{A}/B$ is determined from FUV spectra.} 
\label{prop1}
\centering
\begin{tabular}{ccccccccc}
\hline
\hline
$S_\mathrm{A}/B$&Median $u_\mathrm{r}$& Mean $u_\mathrm{r}$&Median $n_\mathrm{p}$&Mean $n_\mathrm{p}$&Median $P_\mathrm{ram}$&Mean $P_\mathrm{ram}$\\
$\times$10$^6$~W~m$^{-2}$~T$^{-1}$&km~s$^{-1}$&km~s$^{-1}$&cm$^{-3}$&cm$^{-3}$&nPa&nPa\\
\hline
&&&slow wind&\\
1.2&392&377&13.9&22.6&34.1&38.1\\
2.0&375&362&26.7&47.6&60.2&69.0\\
2.2&372&361&30.9&54.7&69.2&78.1\\
\hline
&&&fast wind&\\
1.2&781&776&2.4&2.5&24.3&24.9\\
2.0&703&701&5.9&5.9&48.4&48.4\\
2.2&692&692&6.9&7.0&60.0&55.8\\
\hline
\end{tabular}
\end{table*}
%%%%%%%
%%%%%%%
\begin{table*}
\caption{Median and mean values of the wind speed, proton density, and ram pressure for the slow and the fast wind 
for simulations where $l_\mathrm{max}$=5, and $S_\mathrm{A}/B$ is determined from FUV spectra.} 
\label{prop2}
\centering
\begin{tabular}{ccccccccc}
\hline
\hline
$S_\mathrm{A}/B$&Median $u_\mathrm{r}$& Mean $u_\mathrm{r}$&Median $n_\mathrm{p}$&Mean $n_\mathrm{p}$&Median $P_\mathrm{ram}$&Mean $P_\mathrm{ram}$\\
$\times$10$^6$~W~m$^{-2}$~T$^{-1}$&km~s$^{-1}$&km~s$^{-1}$&cm$^{-3}$&cm$^{-3}$&nPa&nPa\\
\hline
&&&slow wind\\
1.2&423&406&12.4&17.8&34.8&37.3\\
2.0&426&408&23.7&32.6&62.6&70.3\\
2.2&427&410&26.6&37.5&70.9&80.6\\
\hline
&&&fast wind&\\
1.2&791&778&2.1&2.2&21.2&21.7\\
2.0&706&696&5.3&5.3&42.1&42.1\\
2.2&691&683&6.4&6.3&48.7&48.6\\
\hline
\end{tabular}
\end{table*}

\end{appendix}

\end{document}